\documentclass[preprint, amsmath, amssymb, aps]{revtex4-2}

\usepackage{graphicx}
\usepackage{dcolumn}
\usepackage{bm}

\usepackage[caption=false]{subfig} 
\usepackage{multirow}

\allowdisplaybreaks[1]

\newcommand{\hami}{\mathcal}
\newcommand{\del}{\partial}

\begin{document}
    
\preprint{APS/123-QED}

\title{Asymptotically free and safe quantum gravity scenarios consistent with Hubble, laboratory, and inflation scale physics}

\author{Hiroki Hoshina}
\affiliation{Liberality Research, Matsubara 5-22-6, Setagaya-ku, Tokyo 156-0043 Japan}
\affiliation{Institute of Physics, the University of Tokyo, Komaba, Meguro-ku, Tokyo 153-8902 Japan}
\email{hiroki.hoshina@q-bits.jp}


\begin{abstract}

To find out the possible scenarios for quantum gravity consistent with the observed universe, 
we numerically investigate the non-perturbative renormalization group equations of a general quadratic gravity theory recently derived by Sen, Wetterich and Yamada (\textit{JHEP} 03 (2022) 130). 
As boundary conditions, we impose consistency with the Hubble scale and the laboratory scale experiments, and the Starobinsky model of inflation. 
We find two kinds of trajectories which go to different regime at the trans-Planckian scales: 
i) a trajectory which flows to the asymptotically free regime, and 
ii) a trajectory which flows to the asymptotically safe regime. 
To determine the early-time cosmological scenario, 
an additional observational data from beyond the homogeneous and isotropic space-time is necessary. 

\end{abstract}

\maketitle

\section{Introduction \label{sec:intro}}

What possibly happened at the earliest time of the universe evolution from the cosmological point of view? 
There are strong implications to support the scenario that the universe had likely experienced an accelerated expanding era~\cite{Planck:2018jri}, 
and the quantum gravity theory is expected to play a significant role in revealing the dynamics of early universe. 

One of the promising frameworks to understand the inflation mechanism is the asymptotic safety scenario of quantum gravity, also simply called the asymptotic safety~\cite{Weinberg:1980}. 
This is based on Wilsonian renormalization group techniques. 
The core idea of asymptotic safety is 
that gravity is renormalizable in the sense that the quantum nature of gravity is controlled by a non-Gaussian fixed point: 
the renormalizable theory can be defined if the theory has a non-Gaussian fixed point to which the number of \textit{relevant couplings} is finite. 
Since the discovery of 
a non-Gaussian fixed point of gravity the so-called Reuter fixed point~\cite{Reuter:1996cp,Souma:1999at}, 
further evidence for the scenario has been collected (for a recent review see Ref.~\cite{Bonanno:2020bil} and references therein). 

Solutions of the renormalization group equations are represented in the form of trajectories, and the set of all possible trajectories constitute the renormalization group flow. 
Each point of the renormalization group flow corresponds to an energy scale-dependent effective action, which we denote by $\Gamma_k$. 
This effective action is calculated from the fundamental quantum action by partially integrating out the excitations with momenta larger than the energy scale $k$~\cite{Wetterich:1992yh}. 
Thus, the effective action $\Gamma_k$ well describes physics at the typical energy scale $k$ in the tree-level approximation. 
On the other hand, we have to include sufficient loop corrections of $\Gamma_k$ to describe physics at much lower energy scales. 
In this way, the effective action $\Gamma_k$ is supposed to be connecting the low energy and the high energy physics, including the Hubble and the Planck scale physics. 

The energy scale-dependence of the theory plays an important role in cosmology, because the physical parameters for the gravitational theories are obtained from the observations at different energy scales: 
the cosmological constant is estimated from the Hubble scale studies~\cite{Planck:2018vyg}, the Newton constant from the laboratory experiments~\cite{ParticleDataGroup:2020ssz}, 
and the parameters related to inflationary dynamics from the studies at much higher energy scales~\cite{Planck:2018jri}. 
Then, one expects to find out a trajectory which relates all the parameters at given different energy scales, 
and study quantum gravity at much higher energy scales, keeping the consistency with lower energy scale dynamics.

In Ref.~\cite{Gubitosi:2018gsl}, Gubitosi \textit{et al.} 
studied the renormalization group flow of the (Euclidean) $f(R)$ gravity up to the second order, 
\begin{equation}
 S_k = \int d^4x\sqrt{g}\left(U_k-\frac{F_k}{2}R-\frac{C_k}{2}R^2\right), \label{eq:action_GORS}
\end{equation}
where $S_k$ is the tree-level approximation of gravitational quantum action $\Gamma_k$. 
We denote by $x$ the four dimensional space-time coordinate, 
by $g$ the determinant of the metric tensor $g_{\mu\nu}$, and by $R$ the scalar curvature. 
The action is parameterized by the three energy scale-dependent coupling parameters, $U_k$, $F_k$, and $C_k$. 
They numerically analyzed the renormalization group equations (RGEs) of Eq.~(\ref{eq:action_GORS}) 
and found a trajectory which satisfies the observational constraints 
on the cosmological constant, the Newtonian coupling, and $C_k$, assuming the Starobinsky model of inflation.

In this paper, we extend the work of Gubitosi \textit{et al.}~\cite{Gubitosi:2018gsl} 
by studying 
the RGEs of 
gravitational action that includes 
a term proportional to the ``square'' of Weyl tensor, $C_{\mu\nu\rho\lambda}^2=C_{\mu\nu\rho\lambda}C^{\mu\nu\rho\lambda}$. 
The RGEs for this kind of truncated action was derived non-perturbatively by Sen, Wetterich, and Yamada in Ref.~\cite{Sen:2021ffc}. 
We numerically integrate these RGEs to fill all the boundary conditions 
introduced by Gubitosi \textit{et al.} in Ref.~\cite{Gubitosi:2018gsl} with additional conditions for $D_k$, the coupling of the square of Weyl term. 

We find two types of trajectories which satisfy the above boundary conditions: 
i) one goes to asymptotically free regime at the Planck scale, and ii) one goes to asymptotically safe regime at the above Planck scale. 
We will call these trajectories as the asymptotically free and safe trajectories, respectively. 
To determine the Planck scale physics from our approach, 
one needs an observational constraints on the parameter of the square of Weyl tensor $D_k$ from beyond isotropic and homogeneous cosmological studies. 

The rest of this paper is organized as follows. In Section~\ref{sec:setups}, we introduce the non-perturbative RGEs derived in Ref.~\cite{Sen:2021ffc} and boundary conditions obtained from the observations. 
In Section~\ref{sec:analysis}, basic assumptions of our numerical calculation are explained. 
In Section~\ref{sec:results}, we present our results and Section~\ref{sec:conclusion} contains the conclusion.

\section{Setup \label{sec:setups}}

In this study, we analyze the renormalization group equations for coupling parameters in the following quadratic pure gravitational effective action: 
\begin{equation}
    S_k=\int d^4x\sqrt{g}\left(
        U_k-\frac{F_k}{2}R-\frac{C_k}{2}R^2+\frac{D_k}{2}C_{\mu\nu\rho\sigma}^2+\hami{L}_{\rm GB}
        \right), \label{eq:quadGravAction}
\end{equation}
where $D_k$ is the running coupling parameter of the square of Weyl tensor. 
The contribution from the Gauss-Bonnet term is defined by 
\begin{equation}
    \hami{L}_{\rm GB}=E_k(R^2-4R_{\mu\nu}R^{\mu\nu}+R_{\mu\nu\rho\sigma}R^{\mu\nu\rho\sigma}), 
\end{equation}
where $E_k$ is a coupling parameter for the Gauss-Bonnet term. 

Let us introduce dimensionless coupling parameters $u_k$ and $w_k$ as 
\begin{equation}
    u_k = k^{-4}U_k, \qquad w_k = k^{-2}F_k. 
\end{equation}
In Ref.~\cite{Sen:2021ffc}, Sen, Wetterich, and Yamada non-perturbatively derived the RGEs for the coupling parameters of $u_k$, $w_k$, $C_k$, $D_k$, and $E_k$, 
which describe the energy scale dependence of the gravitational action of Eq.~(\ref{eq:quadGravAction}), written in the form of 
\begin{align}
    \del_t u_k = \beta_u(u, w, C, D), \label{eq:beta_u}\\
    \del_t w_k = \beta_w(u, w, C, D), \label{eq:beta_w}\\
    \del_t C_k = \beta_C(u, w, C, D), \label{eq:beta_C}\\
    \del_t D_k = \beta_D(u, w, C, D), \label{eq:beta_D}\\
    \del_t E_k = \beta_E(u, w, C, D), \label{eq:beta_E}
\end{align}
where $\del_t=k\del_k$ is the dimensionless scale derivative. 
The r.h.s. of Eqs.~(\ref{eq:beta_u})--(\ref{eq:beta_E}) are called beta functions. 
The explicit forms of the RGEs (\ref{eq:beta_u})--(\ref{eq:beta_E}) are listed in Appendix~\ref{sec:RGEs}. 
We note that $E_k$ does not enter in the beta functions of all of the couplings, including the beta function of $E_k$ itself. 
Since the Gauss--Bonnet term does not contribute to dynamics in the four dimensional space-time, we will drop Eq.~(\ref{eq:beta_E}) from our consideration. 

To find out a trajectory which is consistent with the universe evolution, we impose boundary conditions for the coupling parameters based on the observational data. 
In this study, we assume the isotropic Friedmann universe evolution and 
treat the Weyl coupling $D_k$ as a free parameter, since the square of Weyl tensor term does not contribute to the dynamics in the isotropic Friedmann universe. 
For $u_k$, $w_k$, and $C_k$, we use the boundary conditions introduced by Gubitosi \textit{et al.} in Ref.~\cite{Gubitosi:2018gsl}. 
Their constraints are expressed as 
\begin{align}
    \Lambda_k=&\frac{k^2u_k}{2w_k}=4\times 10^{-66}~{\rm eV}^2 \quad (k=k_{\rm Hub}), \label{eq:obsConst_Hub}\\
    G_k=&\frac{1}{16\pi k^2w_k}=6.7\times 10^{-57}~{\rm eV}^{-2} \quad (k=k_{\rm lab}), \label{eq:obsConst_lab}\\
    C_k=&1.0\times10^9 \quad (k=k_{\rm inf}), \label{eq:obsConst_inf}
\end{align}
where $\Lambda_k$ is the running cosmological constant, and $G_k$ is the running Newtonian constant. 

Here, we comment on the above constraints. 
In Eq.~(\ref{eq:obsConst_Hub}), it is assumed that the late-time universe evolution is governed by the 
standard cosmology, i.e., the late-time universe is effectively described by the $2\Lambda-R$ action. 
The cosmological constant $\Lambda_k$ is evaluated at the current Hubble scale, $k_{\rm Hub}=10^{-33}~{\rm eV}$. 

In Eq.~(\ref{eq:obsConst_lab}), we impose the condition from the laboratory-scale experiments for the Newtonian coupling parameter. 
The characteristic length scale of the experiments is about $10^{-1}~{\rm m}$, which corresponds to $k_{\rm lab}=10^{-5}~{\rm eV}$. 

In Eq.~(\ref{eq:obsConst_inf}), we presume that the Starobinsky model of inflation~\cite{Starobinsky:1980,Mukhanov:1981xt,Starobinsky:1983zz} effectively describes the early-time universe evolution. 
Since the effective action of the Starobinsky model takes the form of $-R+BR^2$, the running cosmological constant is assumed to be subdominant at $k_{\rm inf}=10^{22}~{\rm eV}$, the approximate Hubble length scale at the inflationary era of the Starobinsky model~\cite{Gubitosi:2018gsl}.

\section{Basic Assumptions \label{sec:analysis}}

Notice that the observational constraints Eqs.~(\ref{eq:obsConst_Hub})--(\ref{eq:obsConst_inf}) are given at different energy scales. 
To find out a trajectory which is consistent with all of these constraints, in our numerical study, we shall obtain a set of initial conditions 
to fill all of these conditions. 
In the following discussion, a relation $A\simeq B$ for given physical values $A$ and $B$ holds if and only if $|(A-B)/A|\ll 1$ holds. 

Since quantum effects of gravity would be negligible at much lower energy scales than the Planck scale, we assume that the running of $C_k$ and $D_k$ are negligible below the inflationary scale: 
\begin{align}
    C_{k_{\rm inf}} &\simeq C_{k_{\rm lab}} \simeq C_{k_{\rm Hub}}, \label{eq:conditions_C}\\
    D_{k_{\rm inf}} &\simeq D_{k_{\rm lab}} \simeq D_{k_{\rm Hub}}. \label{eq:conditions_D}
\end{align}
From Eqs.~(\ref{eq:obsConst_Hub})--(\ref{eq:conditions_D}), we demand 
\begin{align}
    \frac{u_{k_{\rm Hub}}}{w_{k_{\rm Hub}}} &= 8, \label{eq:bound_cond_uw}\\
    w_{k_{\rm lab}} &= 3.0\times 10^{64}, \label{eq:bound_cond_w}\\
    C_{k_{\rm Hub}} &= 1.0\times 10^9 \simeq C_{k_{\rm lab}} \simeq C_{k_{\rm inf}} , \label{eq:bound_cond_C}\\
    D_{k_{\rm Hub}} &\simeq D_{k_{\rm lab}} \simeq D_{k_{\rm inf}}. \label{eq:bound_cond_D}
\end{align}

The numerical integration of RGEs~(\ref{eq:beta_u})--(\ref{eq:beta_D}) can lead to singular effects in the infrared and the ultraviolet regimes. 
The beta functions, the r.h.s. of RGEs~(\ref{eq:beta_u})--(\ref{eq:beta_D}), can be written by the linear combinations of $l^{2n}_p(x)$, which is defined as 
\begin{equation}
    l^{2n}_p(x)=\frac{1}{n!}\frac{1}{(1+x)^{p+1}}, \label{eq:l_2n_p}
\end{equation}
where $n$ and $p$ are integers and $x$ is $\tilde{m}^2_t$ or $\tilde{m}^2_\sigma$, defined as 
\begin{equation}
    \tilde{m}_t^2 = d-v, \qquad \tilde{m}_\sigma^2 = 3c-\frac{v}{4}, 
\end{equation}
where 
\begin{align}
    c = \frac{C}{w}, \qquad d = \frac{D}{w}, \qquad v = \frac{u}{w}. 
\end{align}
Thereby, from Eq.~(\ref{eq:l_2n_p}), the beta functions have a pole at $x=-1$, and we have to check the validity of the RGEs~(\ref{eq:beta_u})--(\ref{eq:beta_D}) while the numerical calculations are performed. 
As discussed in Ref.~\cite{Sen:2021ffc}, the beta functions are valid for $\tilde{m}^2_t>-1$, although more complex for $\tilde{m}^2_\sigma$. 
Yet, we will check whether $\tilde{m}^2_t>-1$ and $\tilde{m}^2_\sigma>-1$ hold or not, and we will take the numerical calculations are valid as far as there is no singular behavior in $\tilde{m}^2_t$, $\tilde{m}^2_\sigma$, and all of the coupling parameters.

\section{Results \label{sec:results}}

We exhibit two types of trajectories which connect the observed universe with i) asymptotically free quantum gravity, and ii) asymptotically safe quantum gravity. 

\subsection{Asymptotically free quantum gravity \label{sec:results_AF}}

Let us assume that the coefficients of the square of scalar curvature and the square of Weyl tensor are of the same order of magnitude at $k_{\rm inf}$. 
By using Eq.~(\ref{eq:conditions_D}), let us set
\begin{equation}
    D_{k_{\rm Hub}} = 1.0\times 10^9. 
\end{equation}
as an input for $D_k$. 
The set of boundary conditions to be satisfied is 
\begin{align}
    \frac{u_{k_{\rm Hub}}}{w_{k_{\rm Hub}}} &= 8, \label{eq:bound_cond_u_AF}\\
    w_{k_{\rm lab}} &= 3.0\times 10^{64}, \label{eq:bound_cond_w_AF}\\
    C_{k_{\rm Hub}} &= 1.0\times 10^9 \simeq C_{k_{\rm lab}} \simeq C_{k_{\rm inf}}, \label{eq:bound_cond_C_AF}\\
    D_{k_{\rm Hub}} &= 1.0\times 10^9 \simeq D_{k_{\rm lab}} \simeq D_{k_{\rm inf}}. \label{eq:bound_cond_D_AF}
\end{align}
We then search the values of $u_{k_{\rm Hub}}$ and $w_{k_{\rm Hub}}$ which realize the above conditions, by numerically integrating out Eqs~(\ref{eq:beta_u})--(\ref{eq:beta_D}). 
The appropriate initial conditions at $k_{\rm Hub}$ are found as 
\begin{align}
    u_{k_{\rm Hub}} &= 4.5\times 10^{121}, \label{eq:ini_cond_u_AF}\\
    w_{k_{\rm Hub}} &= 5.6\times 10^{120}, \label{eq:ini_cond_w_AF}\\
    C_{k_{\rm Hub}} &= 1.0\times 10^9, \label{eq:ini_cond_C_AF}\\
    D_{k_{\rm Hub}} &= 1.0\times 10^9. \label{eq:ini_cond_D_AF}
\end{align}
From these initial conditions, by numerically integrating out the RGEs (\ref{eq:beta_u})--(\ref{eq:beta_D}), we get a trajectory satisfying all the boundary conditions. 

\begin{figure}[htbp]
    \begin{tabular}{c}
        \subfloat[]{
            \includegraphics[width=7.6cm]{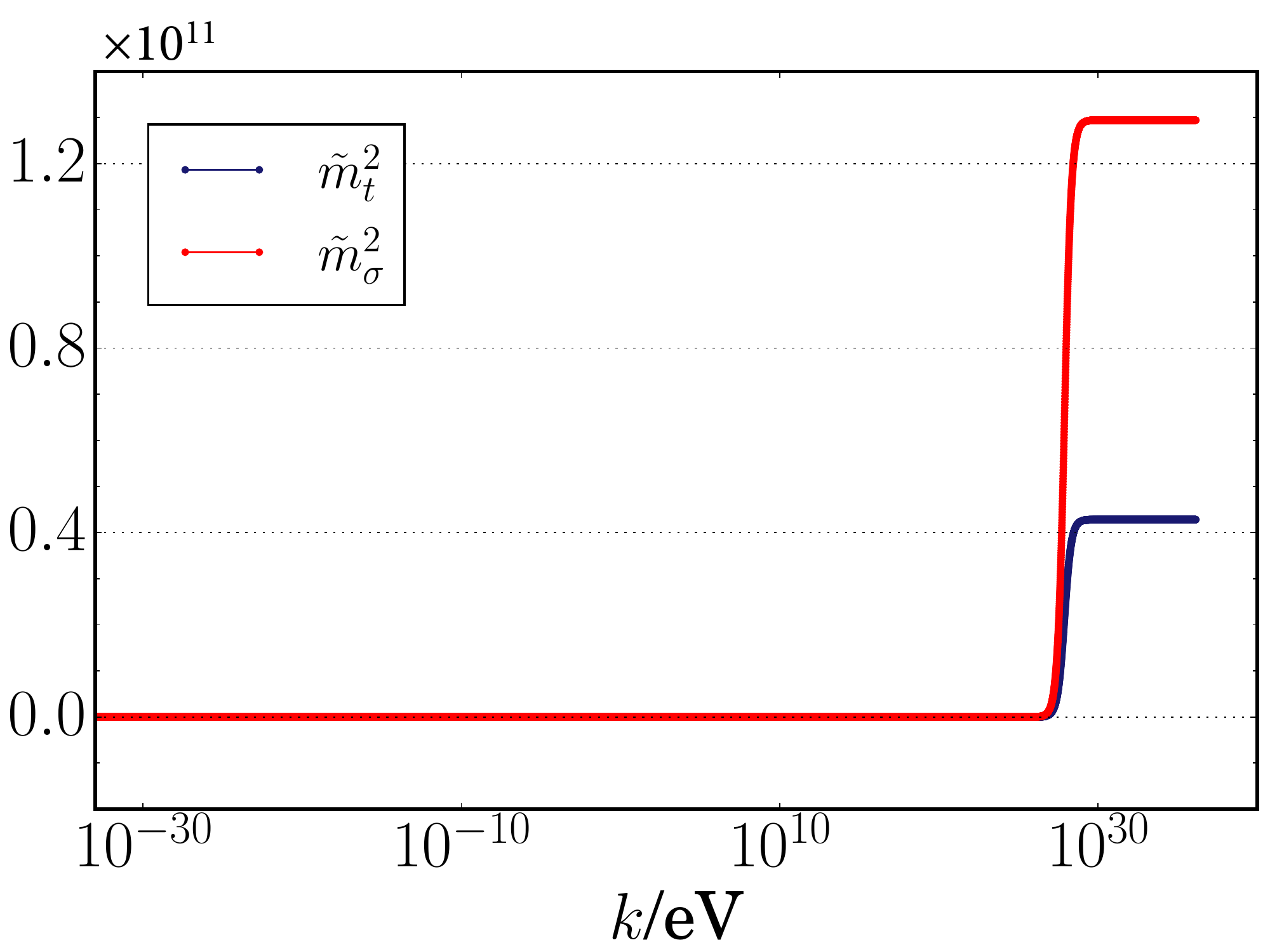}
        } \\
        \subfloat[]{
            \includegraphics[width=7.6cm]{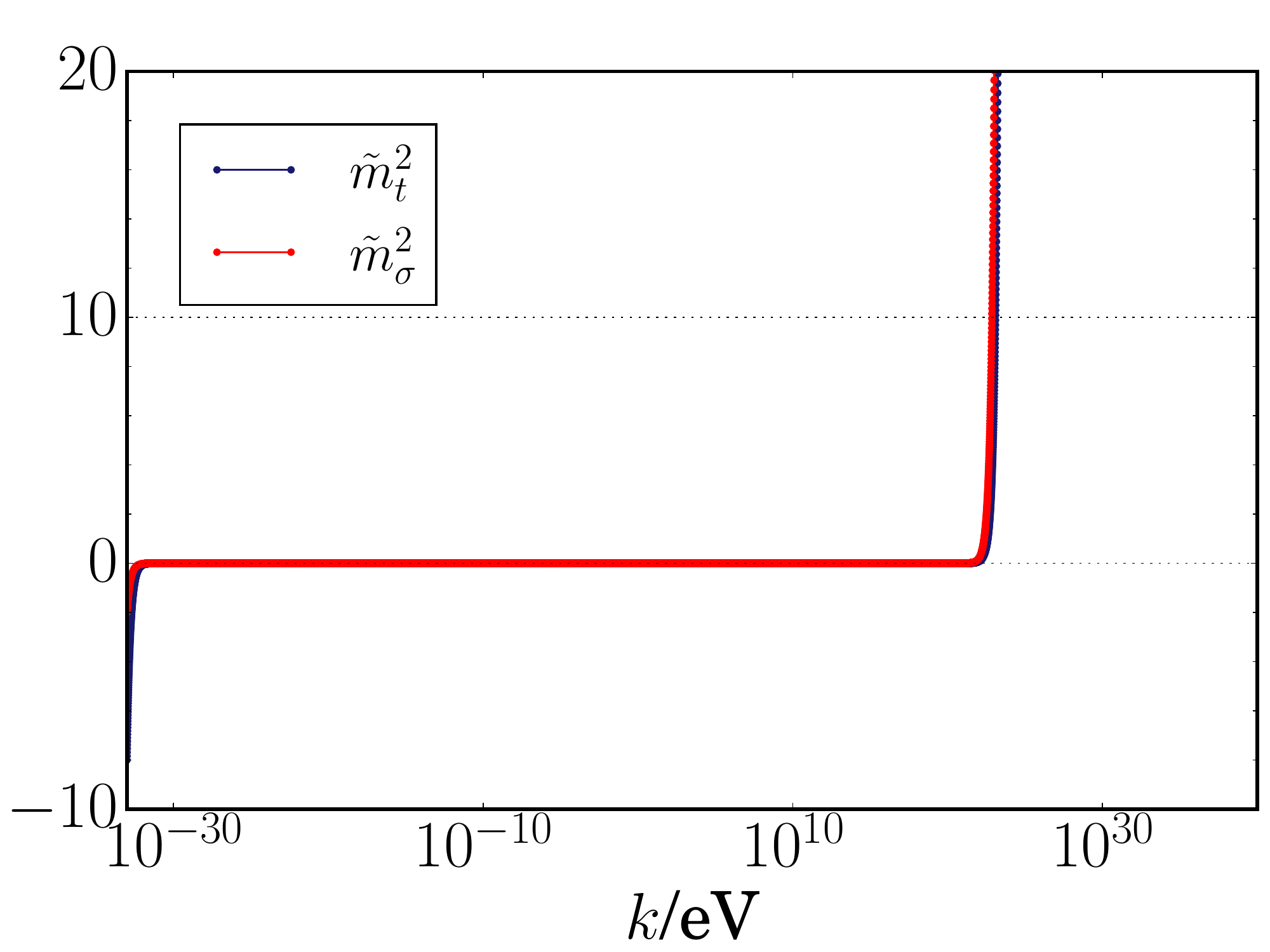}
        } \\
        \subfloat[]{
            \includegraphics[width=7.6cm]{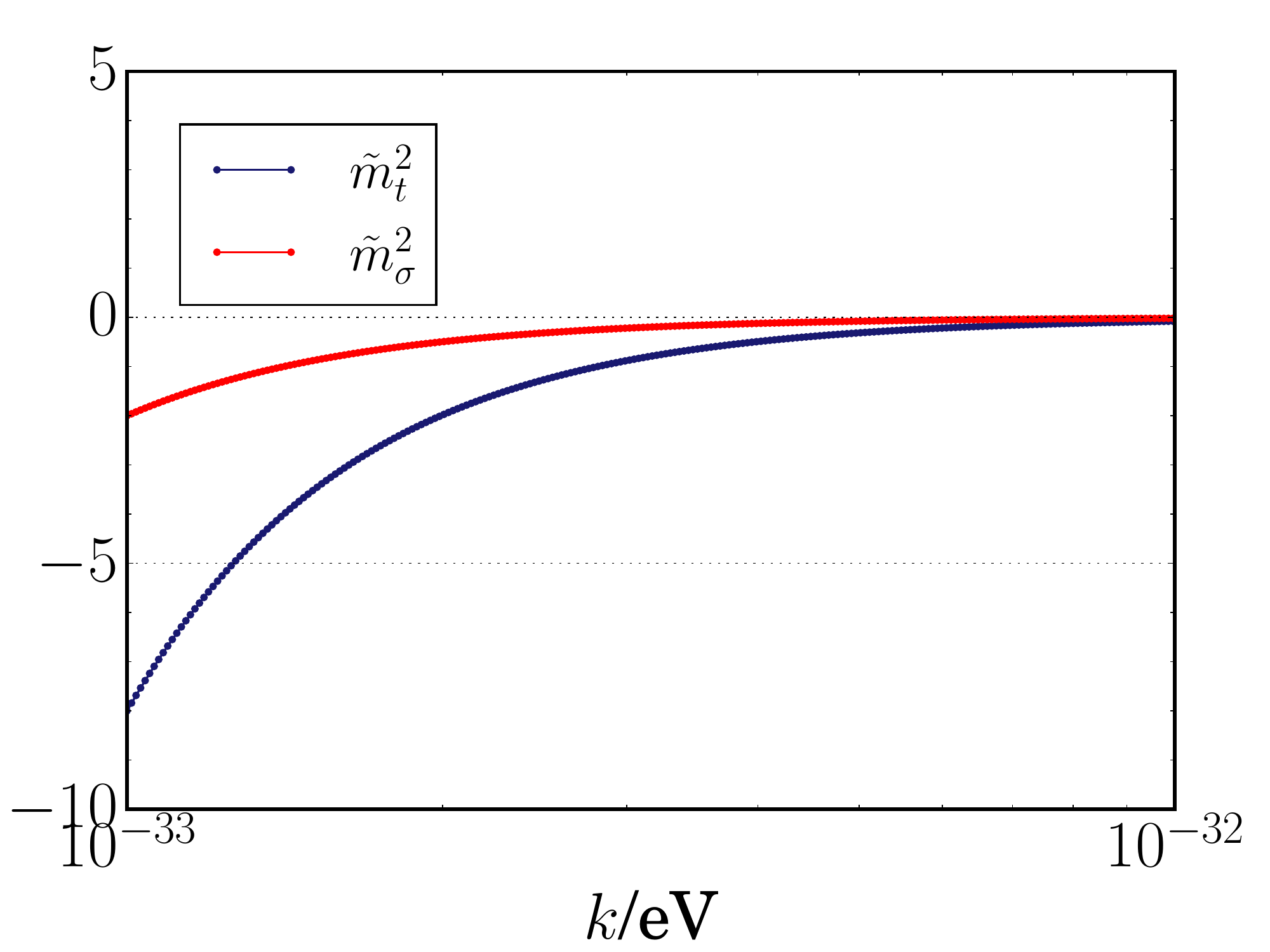}
        }
    \end{tabular}
\caption{The energy scale dependence of $\tilde{m}^2_t$ and $\tilde{m}^2_\sigma$ on the trajectory obtained from the initial conditions of $(u_{k_{\rm Hub}}, w_{k_{\rm Hub}}, C_{k_{\rm Hub}}, D_{k_{\rm Hub}}) = (4.5\times 10^{121}, 5.6\times 10^{120}, 1.0\times 10^9, 1.0\times 10^9)$. 
We can see that $\tilde{m}^2_t$ and $\tilde{m}^2_\sigma$ are continuous in the whole range (top panel) and the validity conditions for the RGEs (\ref{eq:beta_u})--(\ref{eq:beta_D}), $\tilde{m}^2_t, \tilde{m}^2_\sigma > -1$, are fail in the Hubble scale regime (middle and bottom panels). }
\label{fig:SingularitiesAF}
\end{figure}

To see the singularities in the trajectory, we plot the energy scale dependence of $\tilde{m}^2_t$ and $\tilde{m}^2_\sigma$ in Fig.~\ref{fig:SingularitiesAF}. 
Unfortunately, there are regimes near the Hubble scale where $\tilde{m}^2_t, \tilde{m}^2_\sigma > -1$ do not hold. 
However, these singular effects should be unphysical: terms including singularities in the beta functions are quantum corrections (see Eqs.~(\ref{eq:append_beta_u})--(\ref{eq:append_beta_D})), 
and it is unlikely that these quantum effects significantly contributes to the energy scale dependence of coupling parameters. 
In addition, in our calculations, since the contributions of these singular effects to the coupling parameters are $O(10^6)$ and do not contribute to the scale-dependence of the coupling parameters on the trajectory. 
Therefore, in this case, we do not take these contributions from the singularity seriously. 
Note that there is no singular behavior in $\tilde{m}^2_t$ and $\tilde{m}^2_\sigma$. 

The values of coupling parameters at the each energy scale, $k_{\rm Hub}$, $k_{\rm lab}$, and $k_{\rm inf}$, are summarized in Table.~\ref{table:couplings_results_AF}. 
It is easily checked that all of the boundary conditions of Eqs.~(\ref{eq:bound_cond_u_AF})--(\ref{eq:bound_cond_D_AF}) are satisfied. 
\begin{table}[htbp]
 \centering
 \caption{Values of coupling parameters, $u_k$, $w_k$, $C_k$, and $D_k$, at the energy scales of $k_{\rm Hub}$, $k_{\rm lab}$, and $k_{\rm inf}$ on the trajectory obtained from the initial conditions of $(u_{k_{\rm Hub}}, w_{k_{\rm Hub}}, C_{k_{\rm Hub}}, D_{k_{\rm Hub}}) = (4.5\times 10^{121}, 5.6\times 10^{120}, 1.0\times 10^9, 1.0\times 10^9)$. 
 }
 \begin{tabular}{|c|ccc|}
    \hline 
         & $k_{\rm Hub}$ & $k_{\rm lab}$ & $k_{\rm inf}$\\
    \hline
    $u_k$ & $4.5\times10^{121}$ & $1.3\times10^9$ & $4.1\times10^{-3}$ \\ \hline
    $w_k$ & $5.6\times10^{120}$ & $3.0\times10^{64} $& $1.6\times10^{10}$\\ \hline
    $C_k$ & $1.0\times10^9$ & $1.0\times10^9$ & $1.0\times10^{9}$ \\ \hline
    $D_k$ & $1.0\times10^9$ & $1.0\times10^9$ & $1.0\times10^{9}$ \\
    \hline 
 \end{tabular}
 \label{table:couplings_results_AF}
\end{table}

\begin{figure}[htbp]
    \begin{tabular}{cc}
        \subfloat[]{
            \includegraphics[width=0.43\hsize]{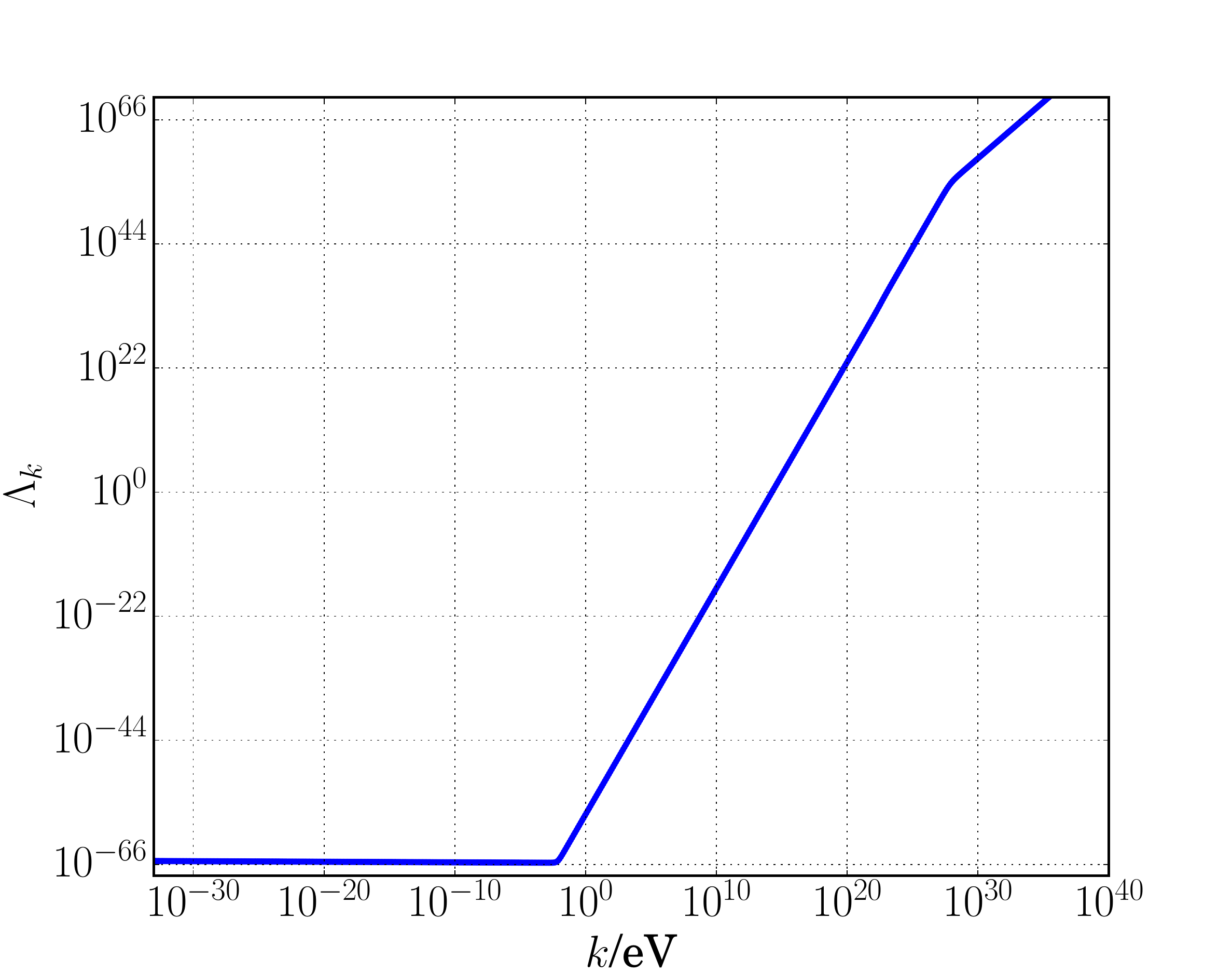}
        } &
        \subfloat[]{
            \includegraphics[width=0.43\hsize]{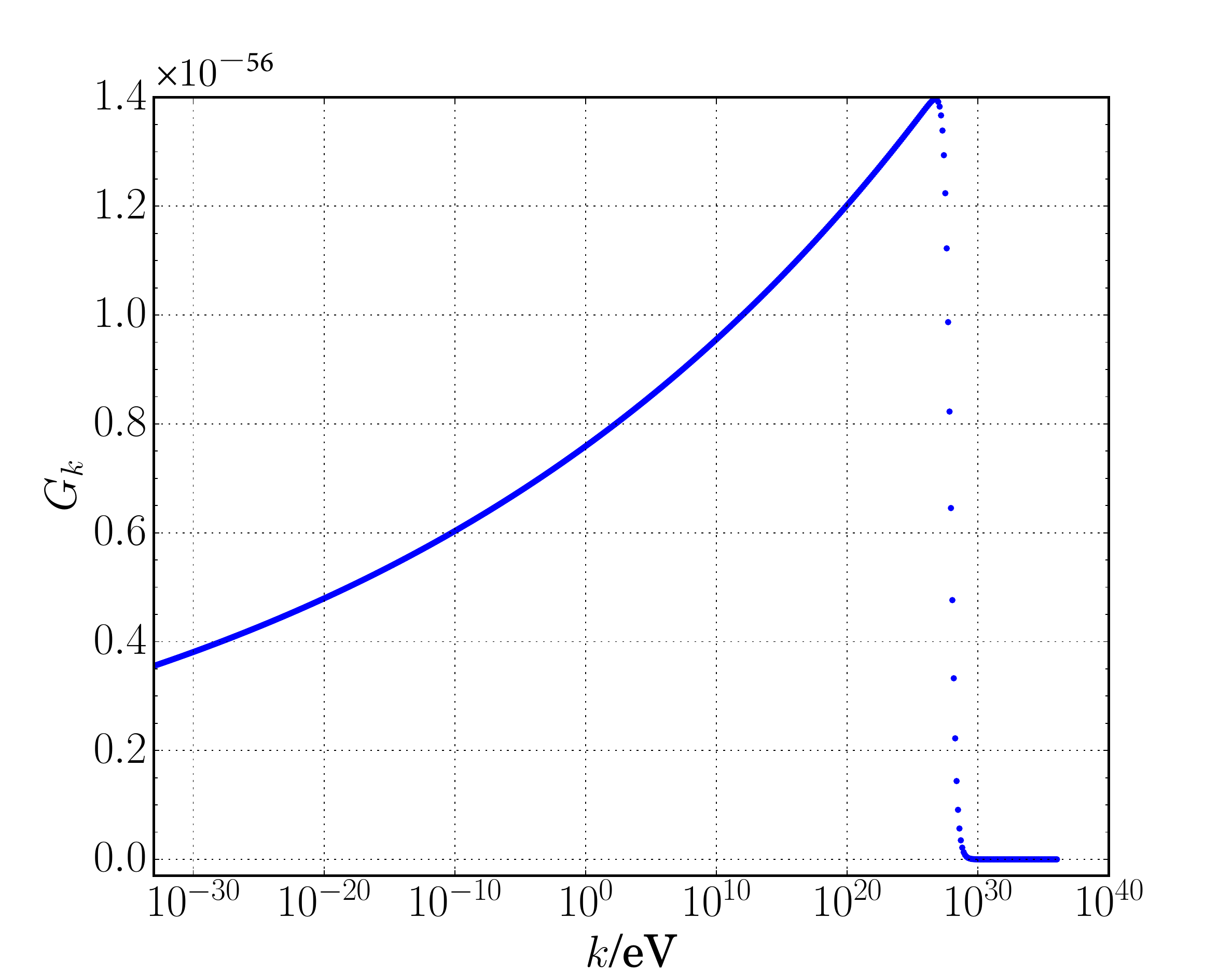}
        }\\
        \subfloat[]{
            \includegraphics[width=0.43\hsize]{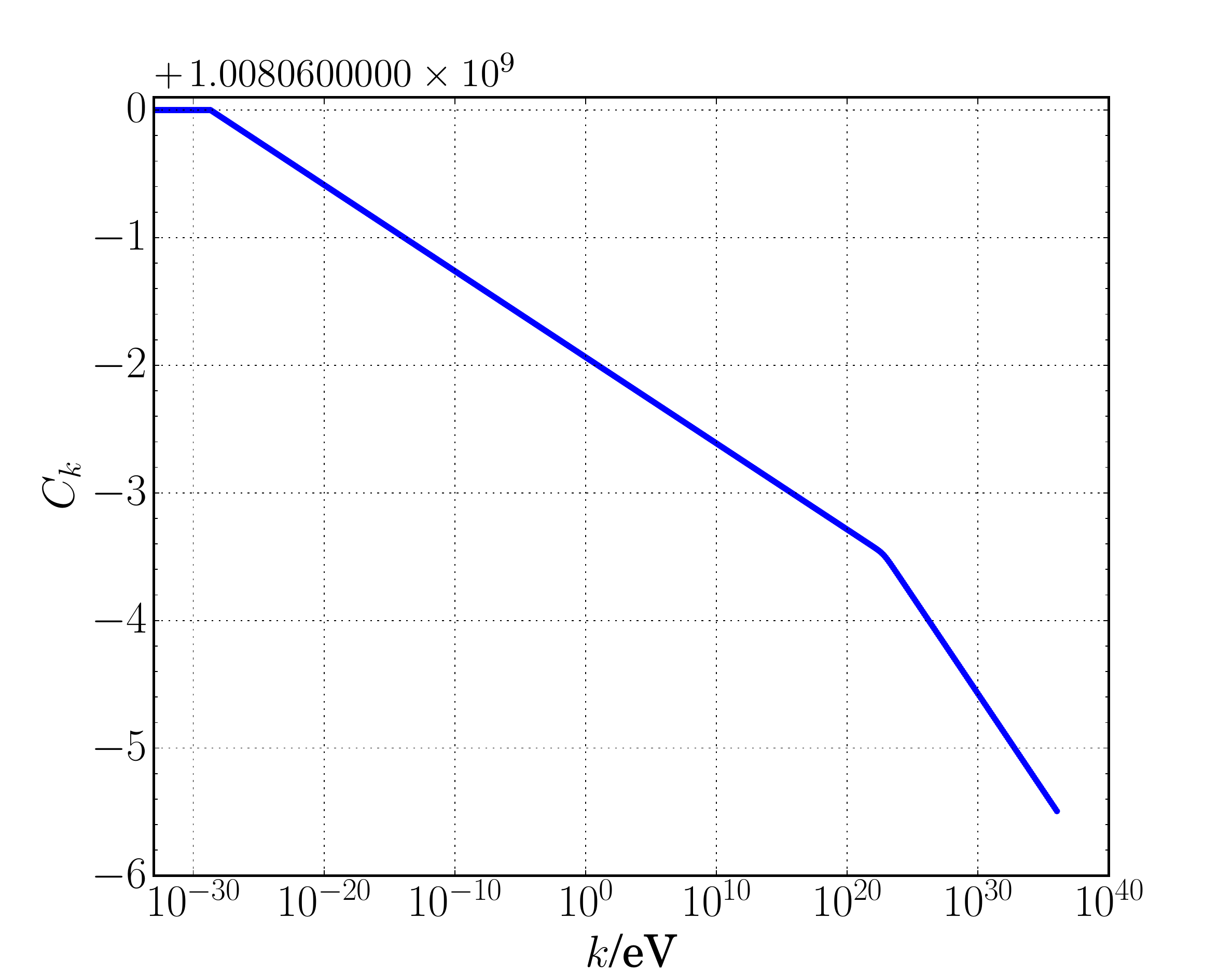}
        } &
        \subfloat[]{
            \includegraphics[width=0.43\hsize]{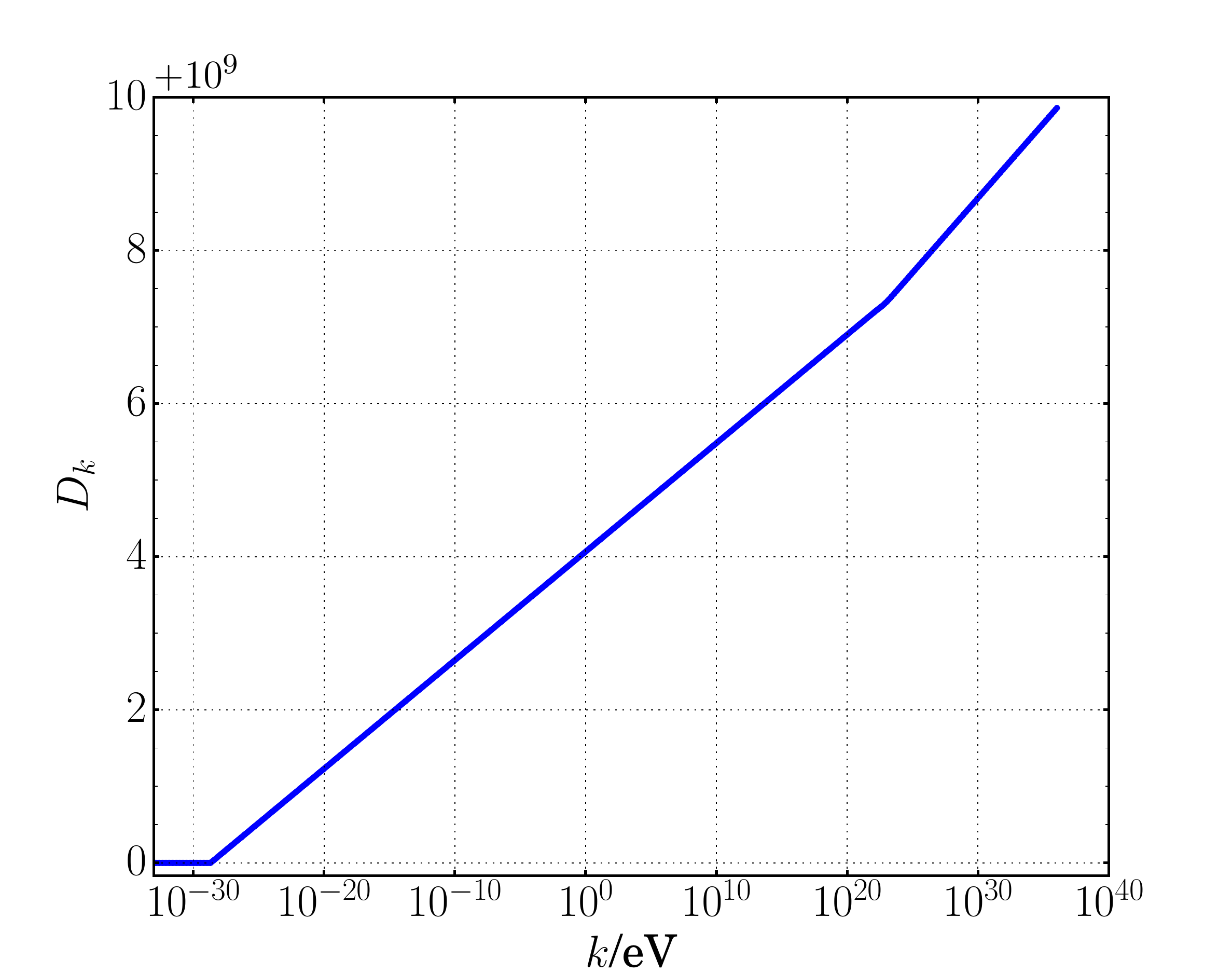}
        }
    \end{tabular}
\caption{The energy scale dependence of the coupling parameters of $\Lambda_k$, $G_k$, $C_k$, and $D_k$ for the initial conditions of 
$(u_{k_{\rm Hub}}, w_{k_{\rm Hub}}, C_{k_{\rm Hub}}, D_{k_{\rm Hub}}) = (4.5\times 10^{121}, 5.6\times 10^{120}, 1.0\times 10^9, 1.0\times 10^9)$. 
}
\label{fig:CouplingsByEnergyScales_AF}
\end{figure}

The energy scale dependence of the coupling parameters of $\Lambda_k$, $G_k$, $C_k$, and $D_k$ are plotted in Fig.~\ref{fig:CouplingsByEnergyScales_AF}. 
There is no singular behavior in all of the coupling parameters. 
We note that the coupling parameters $C_k$ and $D_k$ contribute to the beta functions of Eqs.~(\ref{eq:beta_u})--(\ref{eq:beta_D}) in terms of $C_k/{w_k}$ and $D_k/{w_k}$ (see Appendix \ref{sec:RGEs}). 
In case of the initial conditions of Eqs.~(\ref{eq:ini_cond_u_AF})--(\ref{eq:ini_cond_D_AF}), $C_k/{w_k}$ and $D_k/{w_k}$ are much smaller than $u_k/{w_k}$ and $1$ for $k\lesssim k_{\rm lab}$, 
and thus the contributions from $C_k$ and $D_k$ to the infrared results of $u_k$ and $w_k$ are negligible. 
Therefore, the universal results of the low energy behavior of $\Lambda_k$ and $G_k$ are obtained from such initial conditions. 
We can see that the results of $\Lambda_k$ and $G_k$ are similar to the results of the previous work by Gubitosi \textit{et al.}~\cite{Gubitosi:2018gsl}. 
The slight running of $G_k$ in the infrared regime is supposed to be an artifact of the scheme to derive the RGEs. 

\begin{figure}[htbp]
        \includegraphics[width=12cm]{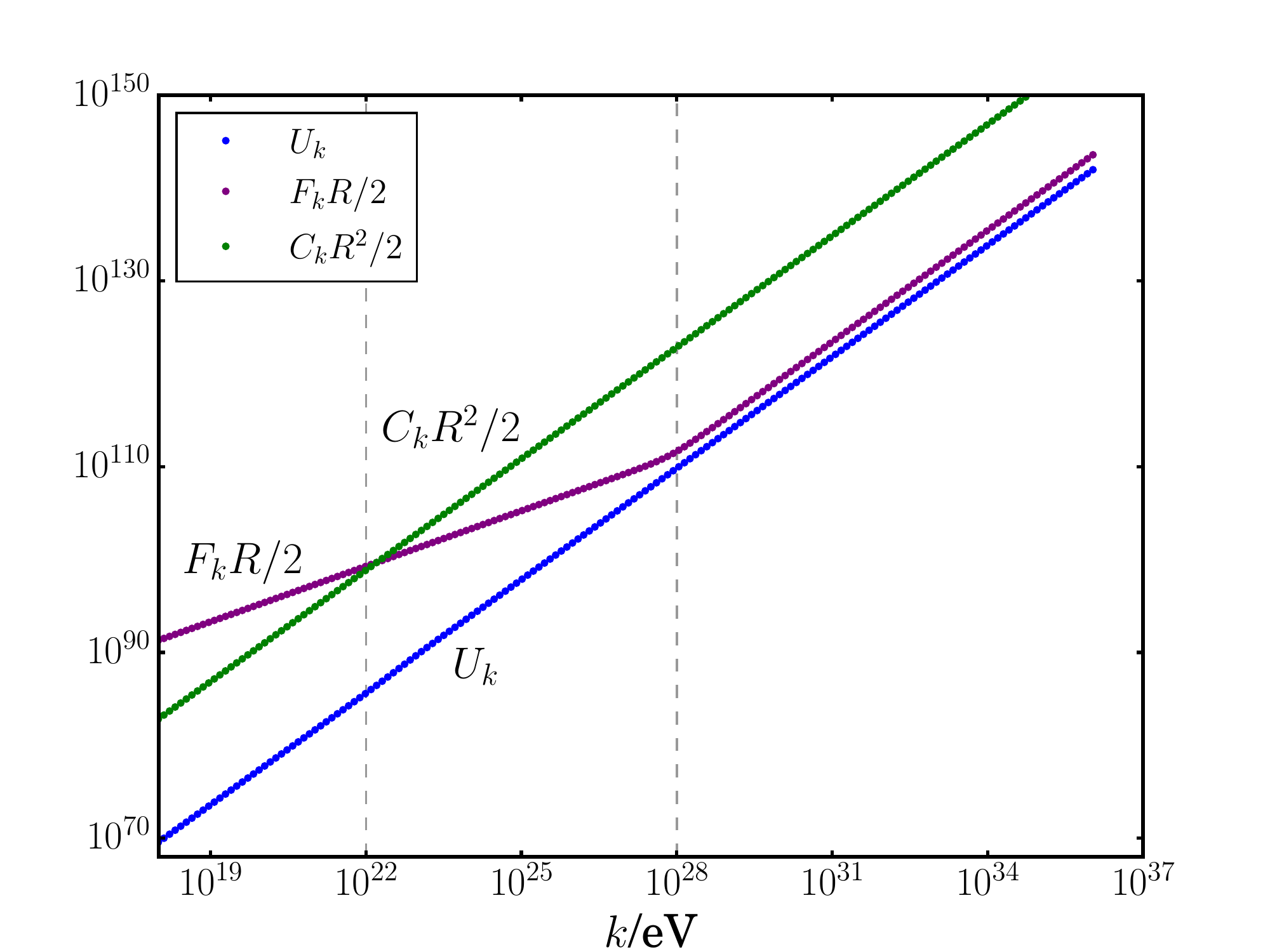}
\caption{Contributions of terms in the action of Eq.~(\ref{eq:quadGravAction}) evaluated in the de Sitter space-time for the initial conditions of 
$(u_{k_{\rm Hub}}, w_{k_{\rm Hub}}, C_{k_{\rm Hub}}, D_{k_{\rm Hub}}) = (4.5\times 10^{121}, 5.6\times 10^{120}, 1.0\times 10^9, 1.0\times 10^9)$. 
}
\label{fig:CompLagrangianTerms_AF}
\end{figure}

Suppose that the universe is approximately described by the de Sitter space-time, at $k\gtrsim k_{\rm inf}$, and compare the contributions of each term in the action of Eq.~(\ref{eq:quadGravAction}). 
Assuming that the Hubble parameter $H$ is comparable with $k_{\rm inf}$, the scalar curvature can be expressed as $R=-12H^2\simeq -12k_{\rm inf}^2$~\cite{Gubitosi:2018gsl}. 
Note that the square of Weyl tensor vanishes in the isotropic Friedmann universe, including the de Sitter space-time. 
However, since $C_k$ and $D_k$ are in the same order at any energy scales of interest, we simply assume that contributions to physics from the quadratic terms, $R^2$ and $C_{\mu\nu\rho\sigma}^2$, are of the same order of magnitude. 
In Fig.~\ref{fig:CompLagrangianTerms_AF}, it is shown that the isotropic Friedmann universe is well described by $-R+R^2$ action (the Starobinsky model) at $k\simeq k_{\rm inf}$, and at higher energy scales, $k\gtrsim k_{\rm pl}$, physics is effectively described by the action containing only the quadratic terms. 
As the coupling values of this theory, $C^{-1}$ and $D^{-1}$, take very small value at $k\gtrsim k_{\rm pl}$, this trajectory corresponds to the asymptotically free quantum gravity theory. 

\subsection{Asymptotically Safe Quantum Gravity \label{sec:results_AS}}

In Ref.~\cite{Sen:2021ffc}, 
a non-trivial fixed point that can be considered as the Reuter fixed point (R-FP)~\cite{Reuter:1996cp,Souma:1999at} has been found: 
\begin{align}
    u^{\rm R}_*=0.000281, \quad w^{\rm R}_*=0.0218, \quad C^{\rm R}_*=0.204, \quad D^{\rm R}_*=-0.0132. 
\end{align}
Quantum physics described at this non-trivial fixed point can be considered as asymptotically safe quantum gravity~\cite{Bonanno:2020bil,Sen:2021ffc}. 
It is very interesting to search for a trajectory which keeps the consistency with the observational conditions and flows into this non-trivial fixed point. 
However, since the critical surface of this non-trivial fixed point is three dimensional~\cite{Sen:2021ffc} 
and the parameter space of the trajectory is four-dimensional, a fine-tuning of the initial conditions is necessary to obtain this trajectory. 
In this paper, instead of performing this task, we study a trajectory which satisfies all of the observational boundary conditions (\ref{eq:bound_cond_uw})--(\ref{eq:bound_cond_D}) and goes to near the R-FP, but run away from the R-FP at the trans-Planckian scales. 
If we set the initial conditions for $u_k$, $w_k$, and $C_k$ as 
\begin{align}
    u_{k_{\rm Hub}} &= 4.4769\times 10^{121}, \label{eq:ini_cond_u_AS}\\
    w_{k_{\rm Hub}} &= 5.5961\times 10^{120}, \label{eq:ini_cond_w_AS}\\
    C_{k_{\rm Hub}} &= 1.0081\times 10^9, \label{eq:ini_cond_C_AS}
\end{align}
the initial condition for $D_k$ that corresponds to such trajectory can be found as 
\begin{equation}
    D_{k_{\rm Hub}} = 535.67. \label{eq:ini_cond_D_AS}
\end{equation}
Here, we have set five significant figures for the coupling parameters. 
The set of the boundary conditions is 
\begin{align}
    \frac{u_{k_{\rm Hub}}}{w_{k_{\rm Hub}}} &= 8, \label{eq:bound_cond_uw_AS}\\
    w_{k_{\rm lab}} &= 3.0\times 10^{64}, \label{eq:bound_cond_w_AS}\\
    C_{k_{\rm Hub}} &= 1.0\times 10^9 \simeq C_{k_{\rm lab}} \simeq C_{k_{\rm inf}} , \label{eq:bound_cond_C_AS}\\
    D_{k_{\rm Hub}} &= 535.67 \simeq D_{k_{\rm lab}} \simeq D_{k_{\rm inf}}. \label{eq:bound_cond_D_AS_1}
\end{align}

\begin{figure}[htbp]
    \begin{tabular}{c}
        \subfloat[]{
            \includegraphics[width=7.6cm]{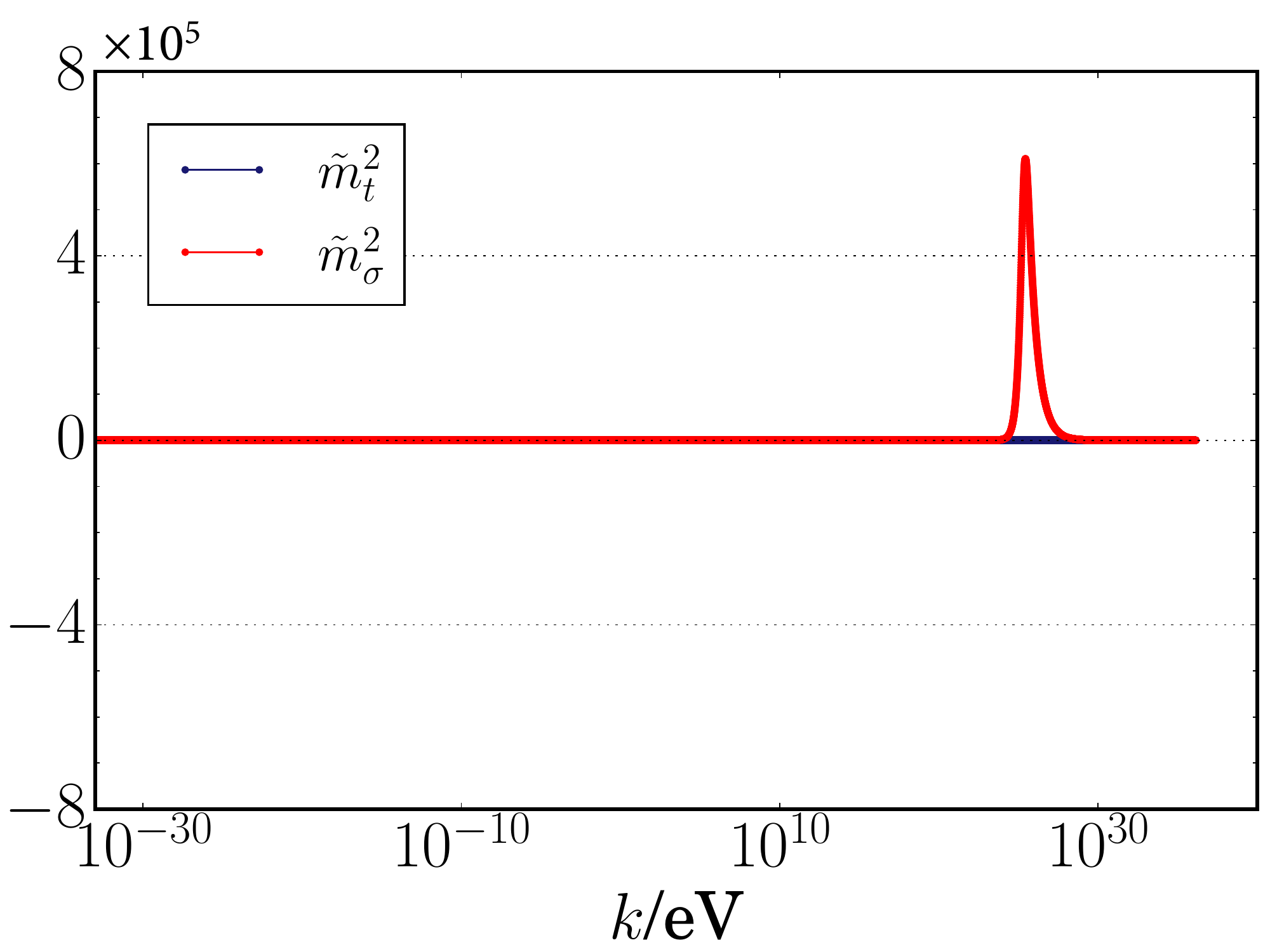}
        } \\
        \subfloat[]{
            \includegraphics[width=7.6cm]{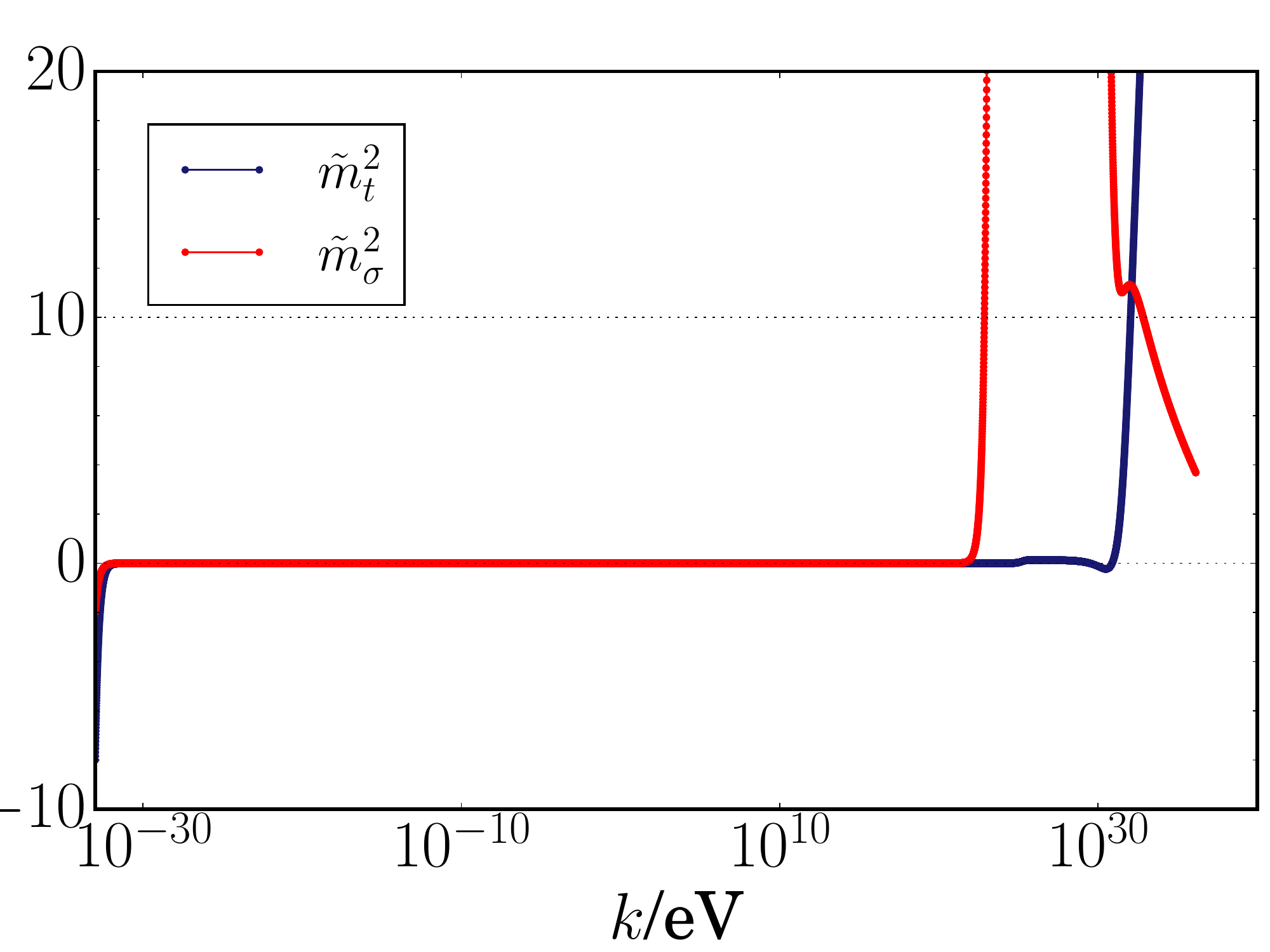}
        } \\
        \subfloat[]{
            \includegraphics[width=7.6cm]{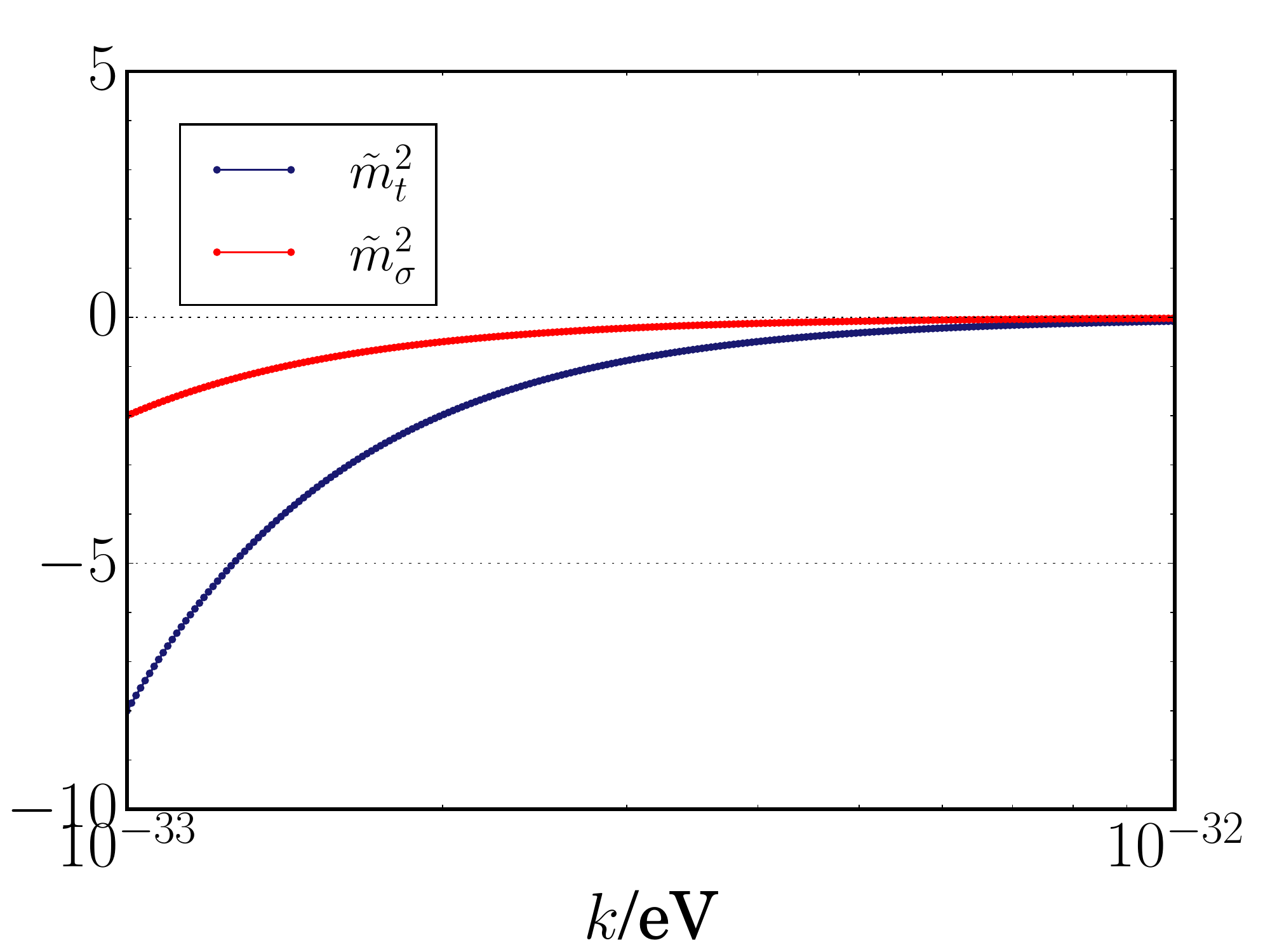}
        }
    \end{tabular}
\caption{The energy scale dependence of $\tilde{m}^2_t$ and $\tilde{m}^2_\sigma$ on the trajectory obtained from the initial conditions of Eqs.~(\ref{eq:ini_cond_u_AS})--(\ref{eq:ini_cond_D_AS}). 
We can see that $\tilde{m}^2_t$ and $\tilde{m}^2_\sigma$ are continuous in the whole range (top panel), and the validity conditions for the RGEs (\ref{eq:beta_u})--(\ref{eq:beta_D}), $\tilde{m}^2_t, \tilde{m}^2_\sigma > -1$, are fail in the Hubble scale regime (middle and bottom panels). 
}
\label{fig:SingularitiesAS}
\end{figure}

The numerical integration of RGEs~(\ref{eq:beta_u})--(\ref{eq:beta_D}) in the initial conditions of Eqs.~(\ref{eq:ini_cond_u_AS})--(\ref{eq:ini_cond_D_AS}) leads to singular effects in the infrared regime. 
As is in the asymptotically free case, there are regimes near the Hubble scale 
where $\tilde{m}^2_t, \tilde{m}^2_\sigma > -1$ do not hold 
(see Fig.~\ref{fig:SingularitiesAS}). 
Since the contributions of the singular effects to the coupling parameters are $O(10^6)$, 
which is much larger than $D_{k_{\rm Hub}}$, the numerical calculations are not reliable. 
To manage the infrared singularity, we assume that these singular effects are unphysical as discussed in the previous subsection, and to eliminate this effects, we set $C_k=C_{k_{\rm Hub}}$ and $D_k=D_{k_{\rm Hub}}$ for $k\in[k_{\rm Hub}, 2.1169\times10^{-29}~{\rm eV}]$. 

\begin{figure}[htbp]
\begin{tabular}{c}
        \includegraphics[width=12cm]{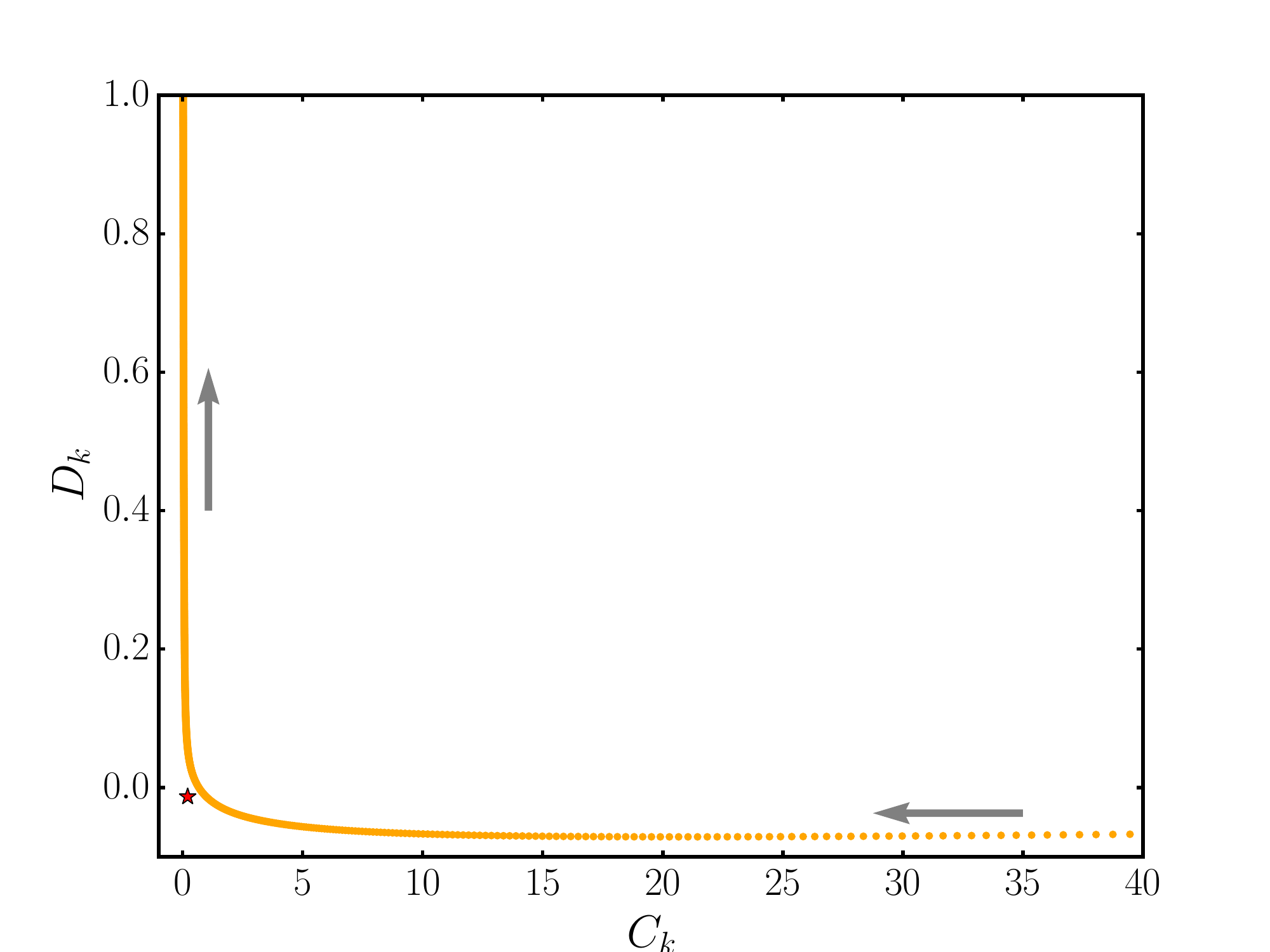}
\end{tabular}
\caption{Renormalization group flow in the $C_k$-$D_k$ plane for the initial condition of Eqs.~(\ref{eq:ini_cond_u_AS})--(\ref{eq:ini_cond_D_AS}). 
The Reuter fixed point, $(u_*^{\rm R}, w_*^{\rm R}, C_*^{\rm R}, D_*^{\rm R})$, is indicated by the star symbol. The gray arrows indicate the direction of increasing energy scale. }
\label{fig:CD_Graph}
\end{figure}

In Fig.~\ref{fig:CD_Graph}, 
we show the renormalization group flow in the $C_k$-$D_k$ plane for the initial conditions of Eqs.~(\ref{eq:ini_cond_u_AS})--(\ref{eq:ini_cond_D_AS}). 
The gray arrows indicate the direction of increasing energy scale. 
It can be seen that the trajectory approaches to the R-FP, indicated by the star symbol, with increasing the energy scale, but $D_k$ begins to increase at near the R-FP. 

The values of coupling parameters at the each energy scale, $k_{\rm Hub}$, $k_{\rm lab}$, and $k_{\rm inf}$, for the initial conditions of Eqs.~(\ref{eq:ini_cond_u_AS})--(\ref{eq:ini_cond_D_AS}) 
are summarized in Table~\ref{table:couplings_results_AS}. 
It can be seen that all the boundary conditions of Eqs.~(\ref{eq:bound_cond_uw_AS})--(\ref{eq:bound_cond_D_AS_1}) are satisfied. 
\begin{table}[htbp]
 \centering
 \caption{Values of coupling parameters, $u_k$, $w_k$, $C_k$, and $D_k$, at the energy scales of $k_{\rm Hub}$, $k_{\rm lab}$, and $k_{\rm inf}$, on the trajectories obtained from the initial conditions of Eqs.~(\ref{eq:ini_cond_u_AS})--(\ref{eq:ini_cond_D_AS}). 
 All the boundary conditions of Eqs.~(\ref{eq:bound_cond_uw_AS})--(\ref{eq:bound_cond_D_AS_1}) are satisfied. 
 }
 \begin{tabular}{|c|ccc|}
    \hline 
         & $k_{\rm Hub}$ & $k_{\rm lab}$ & $k_{\rm inf}$\\
    \hline
    $u_k$ & $4.4769\times10^{121}$ & $1.2567\times10^9$ & $4.0061\times10^{-3}$ \\ \hline
    $w_k$ & $5.5961\times10^{120}$ & $2.9649\times10^{64}$ & $1.5943\times10^{10}$\\ \hline
    $C_k$ & $1.0081\times10^9$ & $1.0081\times10^9$ & $1.0081\times10^{9}$ \\ \hline
    $D_k$ & $535.67$ & $539.03$ & $542.85$ \\
    \hline 
 \end{tabular}
 \label{table:couplings_results_AS}
\end{table}

\begin{figure}[htbp]
    \begin{tabular}{cc}
        \subfloat[]{
            \includegraphics[width=0.43\hsize]{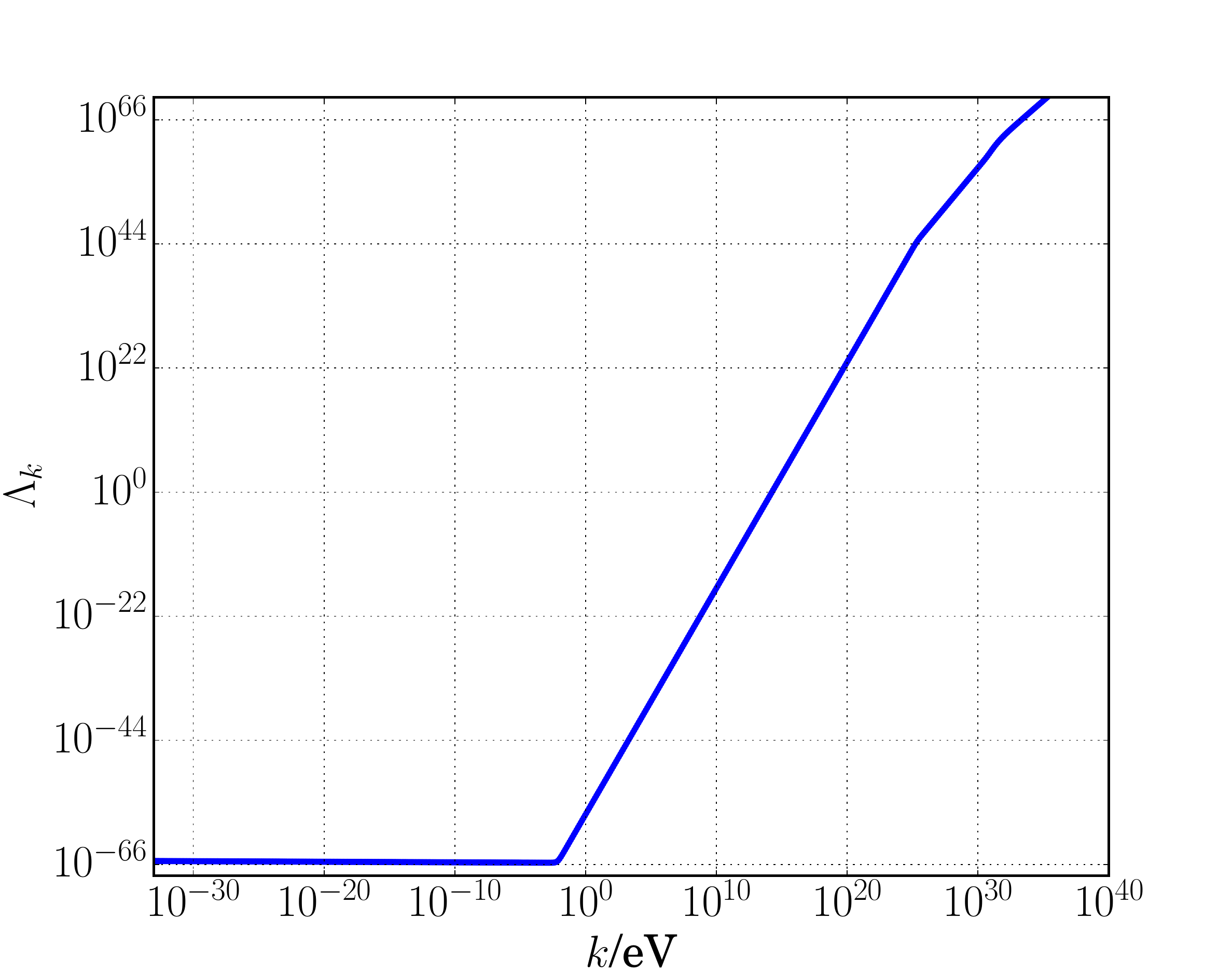}
        } &
        \subfloat[]{
            \includegraphics[width=0.43\hsize]{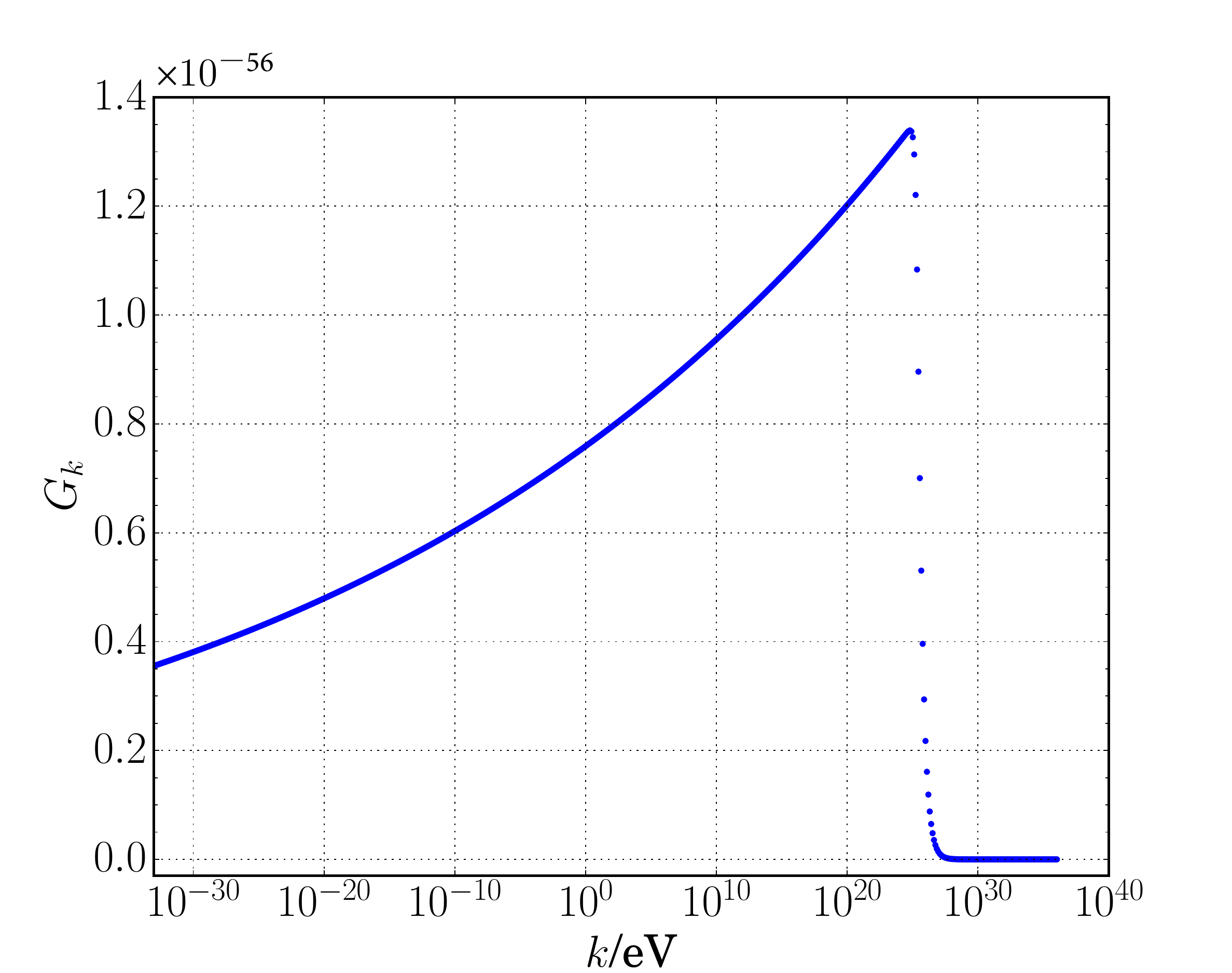}
        }\\
        \subfloat[]{
            \includegraphics[width=0.43\hsize]{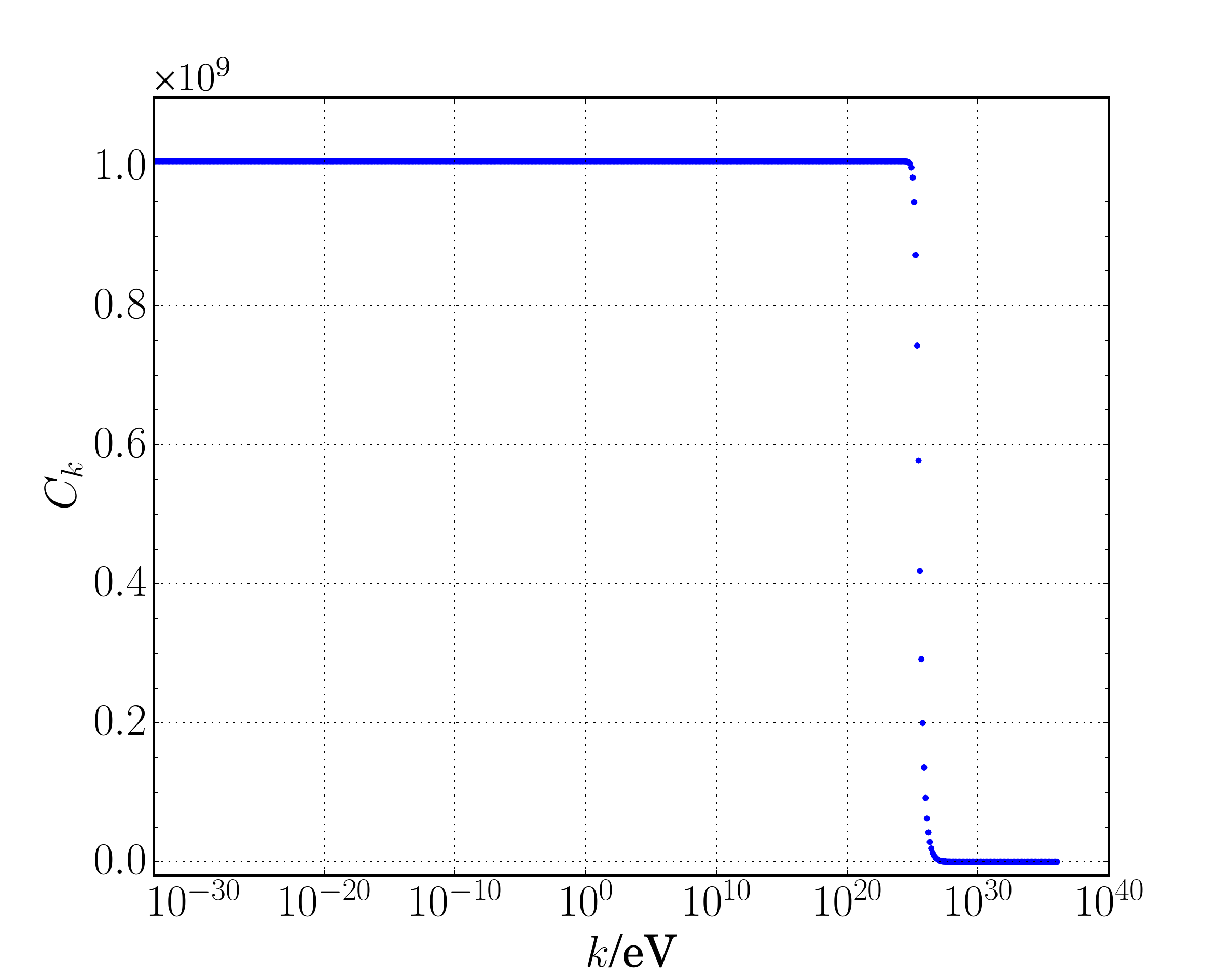}
        } &
        \subfloat[]{
            \includegraphics[width=0.43\hsize]{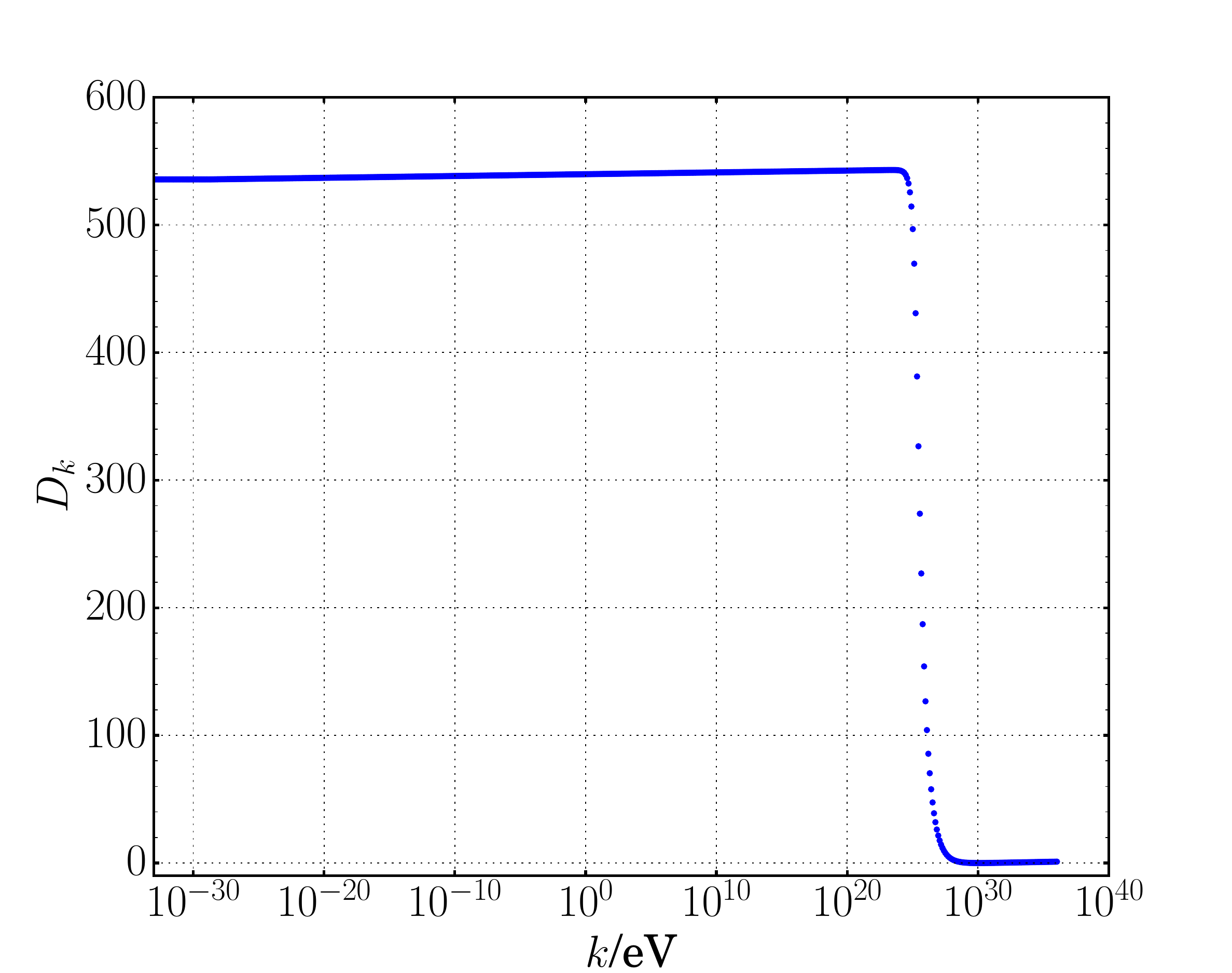}
        }
    \end{tabular}
\caption{The energy scale dependence of the coupling parameters of $\Lambda_k$, $G_k$, $C_k$, and $D_k$ for the boundary condition of 
$(u_{k_{\rm Hub}}, w_{k_{\rm Hub}}, C_{k_{\rm Hub}}, D_{k_{\rm Hub}}) = (4.4769\times 10^{121}, 5.5961\times 10^{120}, 1.008\times 10^9, 535.67)$. 
}
\label{fig:CouplingsByEnergyScales_AS}
\end{figure}

The energy scale dependence of 
the coupling parameters of $\Lambda_k$, $G_k$, $C_k$, and $D_k$ are shown in Fig.~\ref{fig:CouplingsByEnergyScales_AS}. 
As discussed in Section~\ref{sec:results_AF}, since $C_k/w_k$ and $D_k/w_k$ are much smaller than $u_k/w_k$ and $1$ for $k\lesssim k_{\rm lab}$, the results of $\Lambda_k$ and $G_k$ are similar to the asymptotically free case (Fig.~\ref{fig:CouplingsByEnergyScales_AF}) and results of Gubitosi \textit{et al.}~\cite{Gubitosi:2018gsl}. 

\begin{figure}[htbp]
\begin{tabular}{c}
        \includegraphics[width=12cm]{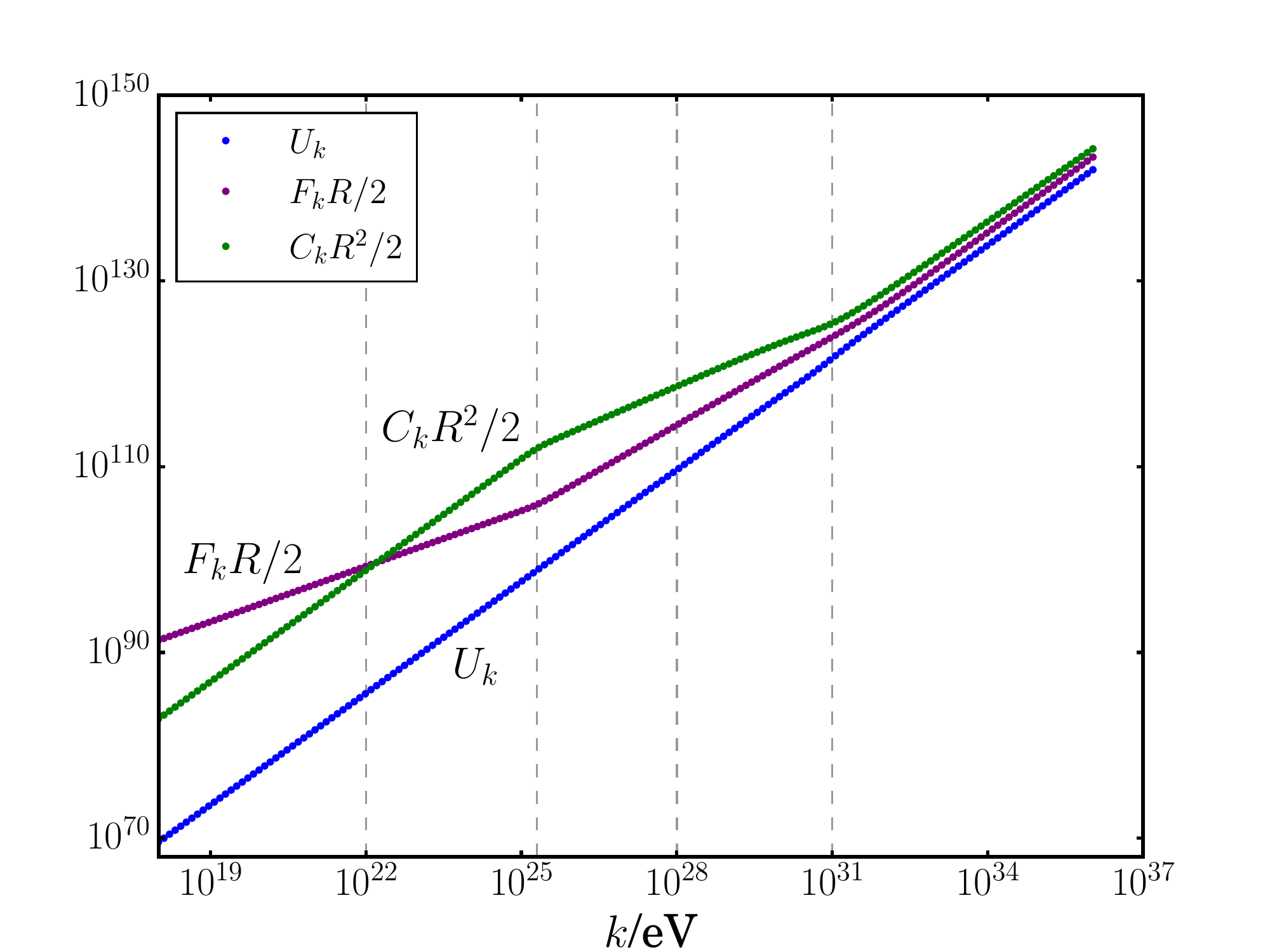}
\end{tabular}
\caption{Contributions of terms in the action of Eq.~(\ref{eq:quadGravAction}) evaluated in the de Sitter space-time for the initial conditions of 
$(u_{k_{\rm Hub}}, w_{k_{\rm Hub}}, C_{k_{\rm Hub}}, D_{k_{\rm Hub}}) = (4.4769\times 10^{121}, 5.5961\times 10^{120}, 1.008\times 10^9, 535.67)$. }
\label{fig:CompLagrangianTerms_AS}
\end{figure}

Contributions of terms in the action in the de Sitter space-time at $k\gtrsim k_{\rm inf}$ are compared in Fig.~\ref{fig:CompLagrangianTerms_AS}. 
Although the theory is well approximated by the Starobinsky model at the inflationary scales, $k\simeq k_{\rm inf}$, the value of $C_kR^2/2$ starts to approach the values of $U_k$ and $F_kR/2$ at $2\times10^{25}~{\rm eV}$, 
and they become of the same order at $10^{31}~{\rm eV}$. 
Although this trajectory does not hit the R-FP, since $C_k$ does not significantly run away from the fixed point value, Fig.~\ref{fig:CompLagrangianTerms_AS} captures the trans-Planckian physics described by the R-FP. 

\section{Conclusion \label{sec:conclusion}}

Keeping the consistency with infrared physics and the Starobinsky model, 
we have found two types of trajectories flowing into 
i) asymptotically free regime and ii) asymptotically safe regime. 
These two trajectories predict the same physics under the inflationary scale, with the assumption that the universe is isotropic Friedmann universe. 
Although the square of Weyl tensor term does not contribute to the homogeneous and isotropic universe dynamics, the boundary condition of $D_k$ has a key roll for the prediction of the Planck scale physics of gravity in the asymptotically safe quantum gravity. 
Thus, the determination of the boundary condition of $D_k$ from cosmological study is important.

\begin{acknowledgments}
I am grateful to Y. Kikukawa for many beneficial discussions. 
I also thank M. Yamada for helpful comments and Y. Narita for improving the manuscript. 
\end{acknowledgments}

\appendix
\section{Renormalization group equations~\label{sec:RGEs}}

To see the explicit forms of renormalization group equations (\ref{eq:beta_u})--(\ref{eq:beta_E}), let us introduce the dimensionless mass terms defined as 
\begin{equation}
    \tilde{m}_t^2 = d-v, \qquad \tilde{m}_\sigma^2 = 3c-\frac{v}{4}, 
\end{equation}
where 
\begin{align}
    c = \frac{C}{w}, \qquad d = \frac{D}{w}, \qquad v = \frac{u}{w}, 
\end{align}
and ``threshold functions'' defined as 
\begin{equation}
    l^{2n}_p(x) = \frac{1}{n!}\frac{1}{(1+x)^{p+1}}, 
\end{equation}
where $n$ and $p$ are integers, and $x$ is a real number. 
Then, the RGEs of Eqs.~(\ref{eq:beta_u})--(\ref{eq:beta_E}) are given as follows. 
Readers interested in details of derivations are referred to the original paper~\cite{Sen:2021ffc}. 

\begin{align}
    \del_t u_k =& -4u+\frac{1}{32\pi^2}\left\{
        \frac{20}{3}(3d+2)l^4_0(\tilde{m}^2_t)+\frac{3}{10}(24+80c-5v)l^4_0(\tilde{m}^2_\sigma)
        -\frac{13}{2}l^4_0(0)
    \right\}, \label{eq:append_beta_u}\\
    \del_t w_k =& -2w-\frac{1}{96\pi^2}\left\{
        -\left[
            \frac{5}{2}(4d+3)l^2_0(\tilde{m}^2_t)+\frac{40}{3}(3d+2)l^4_1(\tilde{m}^2_t) \right.\right. \\
            & \left. +\frac{15}{2}(2c+d)(5+8d)l^6_1(\tilde{m}^2_t)
        \right]
        +
        \left[
            \frac{1}{12}(144c-10v+45)l^2_0(\tilde{m}^2_\sigma) \right. \nonumber\\
            & \left.-\frac{v}{20}(120c-7v+35)l^6_1(\tilde{m}^2_\sigma)+\frac{1}{7}(126c-7v+36)l^8_1(\tilde{m}^2_\sigma) \right. \nonumber\\
            &\left.\left.+\frac{27c}{7}(224c-12v+63)l^{10}_1(\tilde{m}^2_\sigma)-\frac{5}{3}l^2_0(0)
        \right]
        -\frac{29}{4}l^2_0(0)
    \right\}, \label{eq:append_beta_w}\\
    \del_t C_k =& -\frac{1}{576\pi^2}\left\{
        \left[
            -\frac{770}{9}(d+1)l^0_0(\tilde{m}^2_t)+\frac{40}{9}(4d+3)l^2_1(\tilde{m}^2_t)+\frac{40}{27}(3d+2)(3c+7d)l^4_1(\tilde{m}^2_t) \right.\right. \nonumber\\
            & \quad +\frac{1120}{9}(3d+2)l^4_2(\tilde{m}^2_t)+\frac{80}{3}(8d+5)(9c+5d)l^6_2(\tilde{m}^2_t) \nonumber\\
            &\quad +432(5d+3)(2c^2+2cd+d^2)l^8_2(\tilde{m}^2_t)+\frac{5d+3}{50(1+\tilde{m}^2_\sigma)}l^8_1(\tilde{m}^2_t)+\frac{d(12d+7)}{15(1+\tilde{m}^2_\sigma)}l^{10}_1(\tilde{m}^2_t) \nonumber\\
            &\quad \left. -\frac{400(7d+4)(d-6c)^2}{21(1+\tilde{m}^2_\sigma)}l^{12}_1(\tilde{m}^2_t)
        \right] \nonumber\\
        &+\left[
            (12c-v+4)l^0_0(\tilde{m}^2_\sigma)-\frac{7}{20}v(80c-5v+24)l^4_1(\tilde{m}^2_\sigma)+\frac{11}{15}(120c-7v+35)l^6_1(\tilde{m}^2_\sigma)\right. \nonumber\\
            &\quad -\frac{8}{63}(183c-10d)(126c-7v+36)l^8_1(\tilde{m}^2_\sigma)+\frac{20}{21}(21c-d)(126c-7v+36)l^8_1(\tilde{m}^2_\sigma) \nonumber\\
            &\quad +\frac{(126c-7v+36)}{210(1+\tilde{m}^2_t)}l^8_1(\tilde{m}^2_\sigma)+\frac{5}{63}v^2(126c-7v+36)l^8_2(\tilde{m}^2_\sigma) \nonumber\\
            &\quad +\frac{80}{3}c(1080c-55v+297)l^{14}_2(\tilde{m}^2_\sigma)+\frac{60480}{11}c^2(220c-11v+60)l^{16}_2(\tilde{m}^2_\sigma) \nonumber\\
            &\quad +\frac{126c-7v+36}{210(1+\tilde{m}^2_t)}l^8_1(\tilde{m}^2_\sigma)+\frac{3d(224c-12v+63)}{140(1+\tilde{m}^2_t)}l^{10}_1(\tilde{m}^2_\sigma) \nonumber\\
            &\quad \left.\left.-\frac{25(d-6c)^2(288c-15v+80)}{9(1+\tilde{m}^2_t)}l^{12}_1(\tilde{m}^2_\sigma)-\frac{17}{18}l^0_0(0)
        \right] -\frac{29}{2}l^0_0
    \right\}, \label{eq:append_beta_C}\\
    \del_t D_k =& \frac{1}{960\pi^2}\left\{
        \left[
            \frac{1030}{9}(d+1)l^0_0(\tilde{m}^2_t)+\frac{500}{9}(4d+3)l^2_1(\tilde{m}^2_t)+\frac{200}{27}(3d+2)(6c-13d)l^4_1(\tilde{m}^2_t) \right.\right. \nonumber\\
            &\quad +\frac{2800}{9}(3d+2)l^4_2(\tilde{m}^2_t)+\frac{1600}{3}d(8d+5)l^6_2(\tilde{m}^2_t)+4800d^2(5d+3)l^8_2(\tilde{m}^2_t) \nonumber\\
            &\quad +\frac{5d+3}{5(1+\tilde{m}^2_\sigma)}l^8_1(\tilde{m}^2_t)+\frac{2d(12d+7)l^{10}_1(\tilde{m}^2_t)}{3(1+\tilde{m}^2_\sigma)} \nonumber\\
            &\quad \left.-\frac{4000(7d+4)(d-6c)^2}{21(1+\tilde{m}^2_\sigma)}l^{12}_1(\tilde{m}^2_t) 
        \right] \nonumber\\
        &+\left[
            (12c-v+4)l^0_0(\tilde{m}^2_\sigma)+\frac{5}{4}v(-80c+5v-24)l^4_1(\tilde{m}^2_\sigma)+\frac{13}{3}(120c-7v+35)l^6_1(\tilde{m}^2_\sigma) \right. \nonumber\\
            &\quad \frac{20}{63}(183c-10d)(126c-7v+36)l^8_1(\tilde{m}^2_\sigma)+\frac{5}{14}v(224c-12v+63)l^{10}_2(\tilde{m}^2_\sigma) \nonumber\\
            &\quad +\frac{25}{36}(288c-15v+80)l^{12}_2(\tilde{m}^2_\sigma)+\frac{5}{63}v^2(126c-7v+36)l^8_2(\tilde{m}^2_\sigma) \nonumber\\
            &\quad +\frac{50}{3}cv(288c-15v+80)l^{12}_2(\tilde{m}^2_\sigma)+\frac{80}{3}c(1080c-55v+297)l^{14}_2(\tilde{m}^2_\sigma) \nonumber\\
            &\quad \frac{60480}{11}c^2(220c-11v+60)l^{16}_2(\tilde{m}^2_\sigma)+\frac{(126c-7v+36)}{21(1+\tilde{m}^2_t)}l^8_1(\tilde{m}^2_\sigma) \nonumber\\
            &\quad +\frac{3d(224c-12v+63)}{14(1+\tilde{m}^2_t)}l^{10}_1(\tilde{m}^2_\sigma) \nonumber\\
            &\quad \left.\left.-\frac{250(d-6c)^2(288c-15v+80)}{9(1+\tilde{m}^2_t)}l^{12}_1(\tilde{m}^2_\sigma)-\frac{341}{18}l^0_0(0)
        \right] 
        +\frac{7}{2}l^0_0(0)
    \right\}, \label{eq:append_beta_D}\\
    \del_t E_k =& -\frac{1}{5760\pi^2}\left\{
        \left[
            \frac{430}{3}(d+1)l^0_0(\tilde{m}^2_t)+\frac{500}{3}(4d+3)l^2_1(\tilde{m}^2_t)+\frac{200}{9}(3d+2)(6c+5d)l^4_1(\tilde{m}^2_t) \right.\right. \nonumber\\
            &\quad +\frac{1600}{3}(3d+2)l^4_2(\tilde{m}^2_t)+400d(8d+5)l^6_2(\tilde{m}^2_t)+6480d^2(5d+3)l^8_2(\tilde{m}^2_t) \nonumber\\
            &\quad +\frac{3(5d+3)}{5(1+\tilde{m}^2_\sigma)}l^8_1(\tilde{m}^2_t)+\frac{2d(12d+7)}{1+\tilde{m}^2_\sigma}l^{10}_1(\tilde{m}^2_t)-\frac{4000(7d+4)(d-6c)^2}{7(1+\tilde{m}^2_\sigma)}l^{12}_1(\tilde{m}^2_t) \nonumber\\
            &\quad \left. -360(d+1)\frac{N}{\xi_E}l^0_0(\tilde{m}^2_t)
        \right] \nonumber\\
        &+\left[
            (12c-v+4)l^0_0(\tilde{m}^2_\sigma)+\frac{15}{4}v(-80c+5v-24)l^4_1(\tilde{m}^2_\sigma)+13(120c-7v+35)l^6_1(\tilde{m}^2_\sigma) \right.\nonumber\\
            &\quad -\frac{20}{21}(183c-10d)(126c-7v+36)l^8_1(\tilde{m}^2_\sigma)+\frac{15}{14}v(224c-12v+63)l^{10}_2(\tilde{m}^2_\sigma) \nonumber\\
            &\quad +\frac{25}{12}(288c-15v+80)l^{12}_2(\tilde{m}^2_\sigma)+\frac{5}{21}v^2(126c-7v+36)l^8_1(\tilde{m}^2_\sigma) \nonumber\\
            &\quad +50cv(288c-15v+80)l^8_1(\tilde{m}^2_\sigma)+\frac{9d(244c-12v+63)}{14(1+\tilde{m}^2_t)}l^{10}_1(\tilde{m}^2_\sigma) \nonumber\\
            &\quad \left.\left. -\frac{250(d-6c)^2(288c-15v+80)}{3(1+\tilde{m}^2_t)}l^{12}_1(\tilde{m}^2_\sigma)-\frac{317}{6}l^0_0(0)
        \right] 
        -\frac{23}{2}l^0_0(0)
    \right\}. \label{eq:append_beta_E}
\end{align}

\bibliography{trajectoryQuadGrav}

\end{document}